\documentclass{llncs}

\usepackage[ruled,vlined,linesnumbered]{algorithm2e}
\usepackage{comment}
\usepackage{graphicx}
\usepackage{verbatim}
\usepackage{color}
\usepackage{indentfirst}
\usepackage{latexsym,amsmath,epsfig,amssymb,graphics}
\usepackage[ruled,vlined,linesnumbered]{algorithm2e}
\usepackage{amsfonts}
\usepackage{amsmath}
\usepackage{url}
\usepackage{array}
\usepackage{booktabs}
\usepackage{multirow}
\usepackage{makeidx}
\usepackage{times}
\renewcommand{\subparagraph}{}
\usepackage[small,compact]{titlesec}

\newcommand{\oomit}[1]{}

\newcommand{\TT}{\mathcal{T}}
\newcommand{\PT}{\mathbf{PT}}
\newcommand{\NN}{\mathbb{N}}
\newcommand{\QQ}{\mathbb{Q}}
\newcommand{\RR}{\mathbb{R}}
\newcommand{\TQ}{\PT ( \QQ )}
\newcommand{\TR}{\PT ( \RR )}
\newcommand{\ETQ}{\PT ( \QQ )^{\Sigma}}
\newcommand{\ETR}{\PT ( \RR )^{\Sigma}}

\newcommand{\xx}{x}
\newcommand{\sosc}{\mathbf{NSOSC}}
\newcommand{\igfch}{\mathbf{IGFCH}}
\newcommand{\igfqc}{\mathbf{IGFQC}}
\newcommand{\igfqce}{\mathbf{IGFQCE}}
\newcommand{\igfqceu}{\mathbf{IGFQCEunmixed}}

\newcommand{\lmi}{\mathbf{LMI}}
\newcommand{\sdp}{\mathbf{SDP}}

\def \xx {\mathbf{x}}
\def \yy {\mathbf{y}}
\def \zz {\mathbf{z}}
\def \aa {\mathbf{a}}
\def \bb {\mathbf{b}}
\def \cc {\mathbf{c}}

\def \XX {\vec{X}}
\def \va {\vec{\alpha}}
\def \vb {\vec{\beta}}

\newcommand{\xxx}{\mathbf{x}}
\newcommand{\ieq}{\mathbf{IEq}}
\newcommand{\eq}{\mathbf{Eq}}
\newcommand{\leqs}{\mathbf{LEq}}

\begin{document}
\frontmatter          

\title{Interpolation synthesis for quadratic polynomial
  inequalities and combination with \textit{EUF}}
\titlerunning{Hamiltonian Mechanics}  
%
\author{\small Ting Gan\inst{1} \and Liyun Dai\inst{1} \and Bican Xia\inst{1} \and Naijun Zhan\inst{2} \and Deepak Kapur\inst{3}
\and Mingshuai Chen \inst{2}
}
\authorrunning{Ting Gan et al.} 
%
\tocauthor{}
\institute{LMAM \& School of Mathematical Sciences, Peking University\\
\email{\{gant,dailiyun,xbc\}@pku.edu.cn},
\and
State Key Lab. of Computer Science, Institute of Software, CAS \\
\email{znj@ios.ac.cn}
\and
Department of Computer Science, University of New Mexico\\
\email{kapur@cs.unm.edu}
}

\maketitle

\begin{abstract}
An algorithm for generating interpolants for formulas which are
conjunctions of quadratic polynomial
inequalities (both strict and nonstrict) is proposed. The
algorithm is based on a key observation that
quadratic polynomial inequalities can be linearized if they are
concave. A generalization of Motzkin's
transposition theorem is proved, which is used to generate
an interpolant  between two mutually contradictory conjunctions
of polynomial inequalities, using semi-definite programming in time complexity
 $\mathcal{O}(n^3+nm))$ with a given threshold, where $n$ is the number of variables
and $m$ is the number of inequalities.
Using the framework proposed by \cite{SSLMCS2008}
for combining
interpolants for a combination of quantifier-free theories which
have their own interpolation algorithms, a combination algorithm is
given for the combined theory of concave quadratic polynomial
inequalities and the equality theory over uninterpreted functions
symbols (\textit{EUF}). The proposed approach is applicable to all existing abstract domains like
\emph{octagon}, \emph{polyhedra}, \emph{ellipsoid} and so on, therefore it can be used to improve
the scalability of existing verification techniques for programs and hybrid systems.
 In addition, we also discuss how to
extend our approach to formulas beyond concave quadratic polynomials
using Gr\"{o}bner basis.
\end{abstract}

.

\section{Introduction}
Interpolants have been popularized by McMillan \cite{mcmillan05} for automatically generating
invariants of programs. Since then, developing efficient algorithms for
generating interpolants for various theories has become an active area
of research; in
particular, methods have been developed for generating
interpolants for Presburger arithmetic (both for integers as well
as for rationals/reals), theory of equality over uninterpreted
symbols as well as their combination. Most of these methods
assume the availability of a refutation proof of $\alpha \land
\beta$ to generate a ``reverse" interpolant of $(\alpha, \beta)$;
calculi have been proposed to label an inference node in a
refutational proof depending upon whether symbols of formulas on
which the inference is applied are purely from $\alpha$ or
$\beta$.  For propositional calculus, there already existed
methods for generating interpolants from resolution proofs
\cite{krajicek97,pudlak97} prior
to McMillan's work, which generate different interpolants from
those done by McMillan's method. This led D'Silva et al \cite{SPWK10} to study
strengths of various interpolants.

In Kapur, Majumdar and Zarba \cite{KMZ06}, an intimate connection between
interpolants and quantifier elimination was established. Using
this connection, existence of quantifier-free as well as
interpolants with quantifiers were shown for a variety of
theories over container data structures. A CEGAR based
approach was generalized for verification of programs over
container data structures using interpolants. Using this connection between
interpolant
generation and quantifier elimination, Kapur \cite{Kapur13}
has shown that interpolants form a lattice ordered using
implication, with the interpolant generated from $\alpha$ being
the bottom of such a lattice and the interpolant generated from
$\beta$ being the top of the lattice.

Nonlinear polynomials inequalities have been found useful to express
invariants for software involving sophisticated number theoretic
functions as well as hybrid systems; an interested reader may see
\cite{ZZKL12,ZZK13} where different controllers involving nonlinear polynomial
inequalities are discussed for some industrial applications.

We propose an algorithm to generate interpolants for
quadratic
polynomial inequalities (including strict inequalities).
Based on the insight that for analyzing the solution space
of concave quadratic polynomial (strict) inequalities, it
suffices to linearize them. We prove
a generalization of Motzkin's transposition theorem
to be applicable for quadratic polynomial inequalities (including
strict as well as nonstrict).  Based on this result, we
prove the existence of interpolants for two mutually
contradictory conjunctions $\alpha, \beta$ of concave quadratic polynomial
inequalities and give an algorithm for computing an interpolant
using semi-definite programming.
The algorithm is recursive with the basis step of the algorithm
relying on an additional condition on concave quadratic
polynomials appearing in nonstrict inequalities
that any nonpositive constant combination of these polynomials is
never a nonzero sum of square polynomial (called $\sosc$).
In this case, an interpolant output by the algorithm is either a
strict inequality or a nonstrict inequality much like in the
linear case.
In case,
this condition is not satisfied by the nonstrict inequalities,
i.e., there is a nonpositive constant combinations of polynomials
appearing as nonstrict inequalities that is a negative of a sum
of squares, 
then new mutually contradictory conjunctions
of concave quadratic polynomials in fewer variables are derived
from the input augmented with the equality relation deduced, and
the algorithm is recursively invoked on the smaller problem. The
output of this algorithm is in general an interpolant that is a
disjunction of conjunction of polynomial nonstrict or strict inequalities.
The $\sosc$ condition can be checked in polynomial time using
semi-definite programming.

We also show how separating terms $t^-, t^+$ can be constructed using common
symbols in $\alpha, \beta$ such that $\alpha \Rightarrow t^- \le x \le t^+$ and
$\beta \Rightarrow t^+ \le y \le t^-$, whenever $(\alpha \land
\beta) \Rightarrow x = y$. Similar to the construction for
interpolants, this construction has the same recursive structure with
 concave quadratic polynomials satisfying NSOSC as the basis
step.
This result enables the use of the
framework proposed in \cite{RS10} based on
hierarchical theories and a combination method for generating
interpolants by Yorsh and Musuvathi, from combining equality
interpolating quantifier-free theories for generating
interpolants for the combined theory of quadratic polynomial
inequalities and theory of uninterpreted symbols.

Obviously, our results are significant in program verification as
all well-known abstract domains, e.g. \emph{octagon}, \emph{polyhedra}, \emph{ellipsoid} and so on,
which are widely used in the verification of programs and hybrid systems, are
 \emph{quadratic} and \emph{concave}. In addition, we also discuss the possibility to
 extend our results to general polynomial formulas by allowing polynomial equalities whose
 polynomials may be neither \emph{concave} nor \emph{quadratic} using Gr\"{o}bner basis.

We develop a combination algorithm for generating
interpolants for the combination of concave quadratic polynomial
inequalities and uninterpreted function symbols.

In \cite{DXZ13},  Dai et al. gave an
algorithm for generating interpolants for conjunctions of
mutually contradictory nonlinear polynomial inequalities
based on the existence of a witness guaranteed by Stengle's
 \textbf{Positivstellensatz} \cite{Stengle} that can be computed using
semi-definite programming.
Their algorithm is incomplete in general but if every variables ranges
over a bounded interval (called Archimedean condition), then
their algorithm is complete. A major limitation of their work is
that formulas $\alpha, \beta$ cannot have uncommon
variables\footnote{See however an expanded version of their paper
  under preparation where they propose heuristics using program
  analysis for eliminating uncommon variables.}.
However, they do not give any
combination algorithm for generating interpolants in the presence
of uninterpreted function symbols appearing in $\alpha, \beta$.

The paper is organized as follows. After discussing some
preliminaries in the next section,
Section 3 defines concave quadratic polynomials, their matrix
representation and their linearization. Section 4
presents the main contribution of the paper.
A generalization of Motzkin's transposition theorem for quadratic
polynomial inequalities is presented. Using this result,  we
prove the existence of interpolants for two mutually
contradictory conjunctions $\alpha, \beta$ of concave quadratic polynomial
inequalities and give an algorithm (Algorithm 2) for computing an interpolant
using semi-definite programming.
Section 5 extends this algorithm to the combined theory of
concave quadratic inequalities and \textit{EUF} using the framework used
in \cite{SSLMCS2008,RS10}.
Implementation and experimental results using the proposed
algorithms are briefly reviewed in Section 6, and
we conclude and discus future work in Section 7.

\section{Preliminaries}

Let $\NN$, $\QQ$ and $\RR$
be  the set of natural, rational and real numbers, respectively.
Let $\RR[\xx]$ be the polynomial ring over $\RR$ with
variables $\xx=(\xx_1,\cdots,\xx_n)$. An atomic polynomial formula
$\varphi$ is of the form $ p(\xx) \diamond 0$, where $p(\xx) \in
\RR[\xx]$, and
$\diamond$ can be any of  $=, >, \ge, \neq$; without any
loss of generality, we can assume $\diamond$ to be any of $>, \ge$.
An arbitrary polynomial formula is
constructed from atomic ones with Boolean connectives and quantifications over real numbers.
Let $\PT ( \RR )$ be a first-order theory of polynomials with
real coefficient, In this paper, we are focusing on
quantifier-free fragment of $\PT(\RR)$.

Later we discuss quantifier-free theory of equality of terms over
uninterpreted function symbols and its combination with
the quantifier-free fragment of $\PT(\RR)$. Let  $\Sigma$ be a set
of (new) function symbols.
Let $\ETR$ be the extension of
the quantifier-free theory with uninterpreted function symbols in $\Sigma$.

For convenience, we use $\bot$ to stand for \emph{false} and
$\top$ for \emph{true} in what follows.

\begin{definition}
  A model $\mathcal{M}=(M,{f_{\mathcal{M}}})$ of $\ETR$ consists of a model $M$ of
  $\TR$ and a function $f_{\mathcal{M}}: \RR^n \rightarrow \RR$ for each $f\in \Sigma$ with arity $n$.
\end{definition}

\oomit{We have already defined the atomic formulas, thus, a formulas can be defined
easily refer to first-order logic. A formula is called closed, or a sentence, if it
has no free variables. A formula or a term is called ground if it has no occurrences
of variables.

Let $\TT$ be a theory (here is $\TQ$, $\TR$,$\ETQ$ or $\ETR$), we define truth,
satisfiability and entail of a first-order formula in the standard way. }

\begin{definition}
  Let $\phi$ and $\psi$ be formulas of a considered theory $\TT$, then
  \begin{itemize}
    \item $\phi$ is \emph{valid} w.r.t. $\TT$, written as $\models_{\TT} \phi$,  iff $\phi$ is
    true in all models of $\TT$;
    \item $\phi$ \emph{entails} $\psi$ w.r.t. $\TT$, written as $\phi \models_{\TT} \psi$,
    iff for any model of $\TT$, if $\psi$ is true in the model, so is $\phi$;
    \item $\phi$ is \emph{satisfiable} w.r.t. $\TT$, iff there exists a model of $\TT$
    such that in which $\phi$ is true; otherwise \emph{unsatisfiable}.
  \end{itemize}
\end{definition}
Note that $\phi$ is unsatisfiable iff $\phi \models_{\TT} \bot$.

Craig showed that given two formulas $\phi$ and $\psi$ in a
first-order theory $\TT$ such that
$\phi \models \psi$, there always exists an \emph{interpolant} $I$ over
the common symbols of $\phi$ and $\psi$ such that $\phi \models
I, I \models \psi$. In the verification literature, this
terminology has been abused following \cite{mcmillan05}, where an
\emph{reverse} interpolant $I$ over the common symbols of $\phi$ and
$\psi$ is defined for $\phi \wedge\psi \models
\bot$ as: $\phi \models I$ and $I \wedge \psi \models
\bot$.

\begin{definition}
  Let $\phi$ and $\psi$ be two formulas in a theory $\TT$ such that
  $\phi \wedge \psi \models_{\TT} \bot$. A formula $I$ said to be
  a \emph{(reverse) interpolant} of $\phi$ and
  $\psi$ if the following conditions hold:
  \begin{enumerate}
  \item[i] $\phi \models_{\TT} I$;
  \item[ii] $I \wedge \psi \models_{\TT} \bot$; and
  \item[iii]  $I$ only contains common symbols and free variables shared by $\phi$ and
  $\psi$.
  \end{enumerate}
\end{definition}

If $\psi$ is closed, then $\phi \models_{\TT} \psi$ iff
$\phi \wedge \neg \psi \models_{\TT} \bot$. Thus, $I$ is an interpolant of
$\phi$ and $\psi$ iff $I$ is a reverse interpolant of $\phi$ and
$\neg \psi$.
In this paper, we just deal with reveres interpolant, and from now
on, we abuse interpolant and reverse interpolant.

\subsection{Motzkin's transposition theorem}

Motzkin's transposition theorem \cite{schrijver98} is one of the
fundamental results about linear inequalities; it also served as
a basis of the interpolant generation algorithm for the
quantifier-free
theory of linear inequalities in \cite{RS10}.
The theorem has several variants as well.
Below we give two of them.

\begin{theorem}[Motzkin's transposition theorem \cite{schrijver98}] \label{motzkin-theorem} Let $A$ and $B$ be matrices and let $\va$ and $\vb$ be
  column vectors. Then there exists a vector $\xx$ with $A\xx \ge \va$ and $B\xx > \vb$, iff
  \begin{align*}
    &{\rm for ~all~ row~ vectors~} \yy,\zz \ge 0: \\
    &~(i) {\rm ~if~} \yy A + \zz B =0 {\rm ~then~} \yy \va + \zz \vb \le 0;\\
    &(ii) {\rm ~if~} \yy A + \zz B =0 {\rm~and~} \zz\neq 0 {\rm ~then~} \yy \va + \zz \vb < 0.
  \end{align*}
\end{theorem}

\begin{corollary} \label{cor:linear}
  Let $A \in \RR^{r \times n}$ and $B \in \RR^{s \times n}$ be matrices and $\va \in \RR^r$ and
  $\vb \in \RR^s$ be column vectors. Denote by $A_i, i=1,\ldots,r$ the $i$th row of $A$ and by
  $B_j, j=1,\ldots,s$ the $j$th row of $B$. Then there does not exist a vector $\xx$ with
  $A\xx \ge \va$ and $B\xx > \vb$, iff there exist real numbers $\lambda_1,\ldots,\lambda_r \ge 0$
  and $\eta_0,\eta_1,\ldots,\eta_s \ge0$ such that
  \begin{align}
    &\sum_{i=1}^{r} \lambda_i (A_i \xx - \alpha_i) + \sum_{j=1}^{s} \eta_j (B_j \xx -\beta_j) + \eta_0 \equiv 0, \label{eq:corMoz1}\\
    &\sum_{j=0}^{s} \eta_j > 0.\label{eq:corMoz2}
  \end{align}
\end{corollary}

\begin{proof}
The ``if" part is obvious. Below we prove the ``only if" part.

  By Theorem \ref{motzkin-theorem}, if
  $A\xx \ge \va$ and $B\xx > \vb$ have no common solution, then
  there exist two row vectors $\yy \in \RR^r$ and $\zz \in \RR^s$ with $\yy \ge 0$ and $\zz\ge 0$
  such that
   \[ (\yy A+\zz B=0 \wedge \yy \va+ \zz \vb > 0) \vee (\yy A+\zz B=0 \wedge \zz \neq 0 \wedge \yy \va+ \zz \vb  \ge 0).\]
  Let $\lambda_i=y_i, i=1,\ldots, r$, $\eta_j = z_j, j=1,\ldots, s$ and $\eta_0 = \yy \va+ \zz \vb$.
  Then it is easy to check that Eqs. (\ref{eq:corMoz1}) and (\ref{eq:corMoz2}) hold. \qed
\end{proof}

\section{Concave quadratic  polynomials and their linearization}

\begin{definition} [Concave Quadratic] \label{quad:concave}
A polynomial $f \in \RR[\xx]$ is called {\em concave quadratic (CQ)}, if the following two conditions hold:
\begin{itemize}
\item[(i)] $f$ has total degree at most $2$, i.e., it has the form
$f = \xx^T A \xx + 2 \va^T \xx + a$, where $A$ is a real symmetric matrix, $\va$ is a column vector and $a \in \RR$ is a constant;
\item[(ii)] the matrix $A$ is negative semi-definite, written as $A \preceq
0$.\footnote{$A$ being negative semi-definite has many equivalent
  characterizations: for every vector $\xx$, $\xx ^T A \xx \le 0$;
  every $k$th minor of $A$ $\le 0$ if $k$ is odd and $\ge 0$
  otherwise; a Hermitian matrix whose eigenvalues are
nonpositive.}
\end{itemize}
\end{definition}

\begin{example}
Let $g_1= -x_1^2+2 x_1 - x_2^2 + 2 x_2 - y^2$, then it can be expressed as
\begin{align*}
  g_1={\left( \begin{matrix}
    &x_1\\
    &x_2\\
    &y
  \end{matrix} \right)}^T
  {\left( \begin{matrix}
    &-1&0&0\\
    &0&-1&0\\
    &0&0&-1
  \end{matrix} \right)}
  {\left( \begin{matrix}
    &x_1\\
    &x_2\\
    &y
  \end{matrix} \right)} +2 {\left( \begin{matrix}
    &1\\
    &1\\
    &0
  \end{matrix} \right)}^T
  {\left( \begin{matrix}
    &x_1\\
    &x_2\\
    &y
  \end{matrix} \right)}.
\end{align*}
The degree of $g_1$ is 2, and the corresponding $A={\left( \begin{matrix}
    &-1&0&0\\
    &0&-1&0\\
    &0&0&-1
  \end{matrix} \right)} \preceq 0$. Thus, $g_1$ is CQ.
\end{example}

It is easy to see that if $f \in \RR[\xx]$ is linear, then $f$ is
CQ because its total degree is $1$ and the
corresponding $A$
is $0$ which is of course negative semi-definite.

A quadratic polynomial can also be represented as an inner
product of matrices (cf. \cite{laurent}), i.e.,
 $ f(\xx)=\left<P,\left( \begin{matrix}
      1 & \xx^T \\
      \xx  & \xx\xx^T
    \end{matrix}
  \right)\right >.$

\subsection{Linearization} \label{linearization}

Consider quadratic polynomials
$f_i$ and $g_j$ ($i=1,\ldots,r$,
$j=1,\ldots,s$),
\begin{align*}
  f_i=\xx^T A_i \xx+2\va_i^T\xx+a_i,\\
  g_j=\xx^T B_j \xx+2\vb_j^T\xx+b_j,
\end{align*}
where $A_i$, $B_j$ are symmetric $n\times n$ matrices,
$\va_i,\vb_j\in \RR^n$, and $a_i,b_j\in \RR$;
let
$P_i:=\left( \begin{matrix}
    a_i & \va_i^T \\
    \va_i  & A_i
  \end{matrix}
\right),~
Q_j:=\left( \begin{matrix}
    b_j & \vb_j^T \\
    \vb_j  & B_j
  \end{matrix}
\right)$
be
$(n+1)\times(n+1)$ matrices, then
\begin{align*}
  f_i(\xx)=\left<P_i,\left( \begin{matrix}
      1 & \xx^T \\
      \xx  & \xx\xx^T
    \end{matrix}
  \right)\right >,~~
  g_j(x)=\left<Q_j,\left( \begin{matrix}
      1 & \xx^T \\
      \xx  & \xx\xx^T
    \end{matrix}
  \right)\right >.
\end{align*}

For CQ polynomials $f_i$s and $g_j$s in which each $A_i \preceq
0$, $B_j \preceq 0$, define
\begin{equation}
  K=\{\xx \in \RR^n \mid   f_1(\xx) \ge0,\ldots,f_r(\xx)\ge0, g_1(\xx)>0,\ldots, g_s(\xx)>0 \}.
  \label{eq:opt}
\end{equation}

Given a quadratic polynomial
$ f(\xx)=\left<P,\left( \begin{matrix}
      1 & \xx^T \\
      \xx  & \xx\xx^T
    \end{matrix}
  \right)\right >$,
its \emph{linearization} is defined as
  $f(\xx)=\left<P,\left( \begin{matrix}
      1 & \xx^T \\
      \xx  & \XX
    \end{matrix}
  \right)\right >$,
where
     $\left( \begin{matrix}
        1 & \xx^T\\
        \xx & \XX
      \end{matrix}
    \right)\succeq 0$.
    \

Let
  \begin{align*}
  \overline{\XX}=(&\XX_{(1,1)},\XX_{(2,1)},\XX_{(2,2)},\ldots,
  \XX_{(k,1)}, \ldots,\XX_{(k,k)}, \ldots, \XX_{(n,1)}, \ldots,\XX_{(n,n)})
  \end{align*}
    be the vector variable
with $\frac{n(n+1)}{2}$ dimensions corresponding to the matrix $\XX$.
Since $\XX$ is a symmetric
  matrix,
  $\left<P,\left( \begin{matrix}
            1 & \xx^T \\
            \xx & \XX
          \end{matrix}
        \right ) \right >$
  is a linear expression in  $\xx,\overline{\XX}$.

Now, let
  \begin{align}
   &K_1 =  \{\xx \mid  \left( \begin{matrix}
        1 & \xx^T\\
        \xx & \XX
      \end{matrix}
    \right)\succeq 0 ,
    \
    \wedge_{i=1}^r \left<P_i,\left( \begin{matrix}
	  1 & \xx^T \\
	  \xx & \XX
	\end{matrix}
      \right ) \right > \ge 0 , \nonumber \\
     & \quad \quad \quad \quad
    \wedge_{j=1}^s \left<Q_j,\left( \begin{matrix}
	  1 & \xx^T \\
	  \xx & \XX
	\end{matrix}
      \right ) \right > > 0,  \mbox{ for some } \XX \},\label{eq:mom1}
  \end{align}
which is the set of all $\xx\in \RR^n$ on linearizations of the above $f_i$s and $g_j$s.


\oomit{Thus,
\begin{aligned}
\left<P_i,\left( \begin{matrix}
	  1 & \xx^T \\
	  \xx & \XX
	\end{matrix}
      \right ) \right > \ge 0 ,
      \
    &\left<Q_j,\left( \begin{matrix}
	  1 & \xx^T \\
	  \xx & \XX
	\end{matrix}
      \right ) \right > > 0,
 \end{aligned}

are respectively linearizations of quadratic nonstrict inequalities $f_i
\ge 0$'s and strict inequalities $g_j > 0$'s in which all
quadratic terms are abstracted by new variables.
A reader should note that the condition

  \begin{aligned}
     \left( \begin{matrix}
        1 & \xx^T\\
        \xx & \XX
      \end{matrix}
    \right)\succeq 0.
\end{aligned}
is critical.
}

In \cite{fujie,laurent}, when $K$ and $K_1$ are defined only with $f_i$ without $g_j$, i.e., only with
non-strict inequalities, it is proved that $K=K_1$.
\oomit{The set $K$ is defined by a set of quadratic inequalities. The
set $K_1$ is defined by a positive semi-definite constraint and a
set of linear inequalities.}
 By the following Theorem \ref{the:1},
we show that  $K=K_1$ also holds even in
the presence of strict inequalities when $f_i$ and $g_j$ are CQ. So, when
$f_i$ and $g_j$ are CQ, the  CQ
 polynomial inequalities can be transformed  equivalently to a set of
linear inequality constraints and a positive semi-definite
constraint.

\begin{theorem} \label{the:1}
  Let $f_1,\ldots,f_r$ and $g_1,\ldots,g_s$ be CQ polynomials, $K$ and
  $K_1$ as above, then $K=K_1$.
\end{theorem}
\begin{proof}
  For any $\xx \in K$, let $\XX=\xx \xx^T$. Then it is easy to see that
  $\xx,\XX$ satisfy (\ref{eq:mom1}). So $\xx \in K_1$, that is $K \subseteq K_1$.

  Next, we prove $K_1 \subseteq K$.
  Let $\xx \in K_1$, then there exists a symmetric $n\times n $ matrix $\XX$ satisfying
  (\ref{eq:mom1}).
 Because
$\left( \begin{matrix}
        1 & \xx^T\\
        \xx & \XX
      \end{matrix}
      \right)\succeq 0$,
we have $\XX - \xx \xx^T \succeq 0$.
Then by the last two conditions in (\ref{eq:mom1}), we have
\begin{align*}
  f_i(x) &= \left< P_i,\left( \begin{matrix}
            1 & \xx^T \\
            \xx & \xx \xx^T
          \end{matrix}
        \right ) \right> =
        \left<P_i,\left( \begin{matrix}
            1 & \xx^T \\
            \xx & \XX
          \end{matrix}
        \right ) \right> +
        \left<P_i,\left( \begin{matrix}
            0 & 0 \\
            0 & \xx \xx^T - \XX
          \end{matrix}
        \right ) \right> \\
        &=\left<P_i,\left( \begin{matrix}
            1 & \xx^T \\
            \xx & \XX
          \end{matrix}
        \right ) \right> +
        \left<A_i , \xx \xx^T - \XX \right> \ge \left<A_i , \xx \xx^T - \XX \right>, \\[1mm]
  g_j(x) &= \left< Q_j,\left( \begin{matrix}
            1 & \xx^T \\
            \xx & \xx \xx^T
          \end{matrix}
        \right ) \right> =
        \left<Q_j,\left( \begin{matrix}
            1 & \xx^T \\
            \xx & \XX
          \end{matrix}
        \right ) \right> +
        \left<Q_j,\left( \begin{matrix}
            0 & 0 \\
            0 & \xx \xx^T - \XX
          \end{matrix}
        \right ) \right> \\
        &=\left<Q_j,\left( \begin{matrix}
            1 & \xx^T \\
            \xx & \XX
          \end{matrix}
        \right ) \right> +
        \left<B_j , \xx \xx^T - \XX \right> > \left<B_j , \xx \xx^T - \XX \right>.
\end{align*}
Since $f_i$ and $g_j$ are all CQ, $A_i \preceq 0$ and $B_j \preceq 0$.
Moreover,  $\XX -\xx \xx^T \succeq 0$, i.e.,
$\xx \xx^T -\XX \preceq 0$. Thus,
$\left<A_i , \xx \xx^T - \XX \right> \ge 0$ and
$\left<B_j , \xx \xx^T - \XX \right> \ge 0$.
Hence, we have $f_i(\xx) \ge 0$ and $g_j(\xx) > 0$, so $\xx \in K$,
  that is
$K_1 \subseteq K$.
  \qed
\end{proof}

\subsection{Motzkin's theorem in Matrix Form}
If
  $\left<P,\left( \begin{matrix}
            1 & \xx^T \\
            \xx & \XX
          \end{matrix}
        \right ) \right >$
  is seen as  a linear expression in $\xx,\overline{\XX}$,  
then Corollary \ref{cor:linear}
can be reformulated as:
\begin{corollary}  \label{cor:matrix}
  Let $\xx$ be a column vector variable of dimension $n$ and $\XX$ be a $n \times n$
  symmetric matrix variable. Suppose $P_0,P_1,\ldots, P_r$ and $Q_1,\ldots, Q_s$ are
   $(n+1) \times (n+1)$ symmetric matrices. Let
  \begin{align*}
    W \hat{=} \{ (\xx,\XX) \mid
    \wedge_{i=1}^r \left<P_i,\left( \begin{matrix}
            1 & \xx^T \\
            \xx & \XX
          \end{matrix}
        \right ) \right > \ge 0,
   \wedge_{i=1}^s  \left<Q_j,\left( \begin{matrix}
            1 & \xx^T \\
            \xx & \XX
          \end{matrix}
        \right ) \right > > 0 \oomit{,
        ~for~
        \begin{matrix}
            i=0,1,\ldots,r, \\
            j=1,\ldots,s.
          \end{matrix} }
    \},
  \end{align*}
then $W=\emptyset$ iff there exist $\lambda_0, \lambda_1,\ldots,\lambda_r \ge 0$
and $\eta_0,\eta_1,\ldots,\eta_s \ge 0$ such that
\begin{align*}
  &\sum_{i=0}^{r}\lambda_i \left<P_i,\left( \begin{matrix}
            1 & \xx^T \\
            \xx & \XX
          \end{matrix}
        \right ) \right > +
  \sum_{j=1}^{s}\eta_j \left<Q_j,\left( \begin{matrix}
            1 & \xx^T \\
            \xx & \XX
          \end{matrix}
        \right ) \right > + \eta_0 \equiv 0, \mbox{ and }\\
  &\eta_0 + \eta_1 + \ldots + \eta_s > 0.
\end{align*}
\end{corollary}


\section{Algorithm for generating interpolants for Concave
  Quadratic Polynomial inequalities}

\begin{problem} \label{CQ-problem}
Given two formulas $\phi$ and $\psi$
 on $n$ variables with $\phi \wedge \psi
\models \bot$, 
where
\begin{eqnarray*}
 \phi & = &  f_1 \ge 0 \wedge \ldots \wedge f_{r_1} \ge 0 \wedge g_1 >0 \wedge \ldots \wedge g_{s_1} > 0, \\
 \psi & =  & f_{r_1+1} \ge 0 \wedge \ldots \wedge f_{r} \ge 0 \wedge g_{s_1+1} >0 \wedge \ldots \wedge g_{s} > 0,
 \end{eqnarray*}
 in which
$f_1, \ldots, f_{r}, g_1, \ldots, g_s$ are all CQ,
develop an algorithm   
  to generate  a (reverse) Craig interpolant
 $I$  for $\phi$ and  $\psi$, on the common variables of
  $\phi$ and $\psi$,
  such that $\phi \models I$ and $I \wedge
  \psi \models \bot$. For convenience, we partition the variables appearing in the
polynomials above into three disjoint subsets $\xx=(x_1,\ldots,x_d)$ to stand for
the common variables appearing in both $\phi$ and $\psi$, $\yy=(y_1,\ldots,y_u)$ to stand for the
variables appearing only in $\phi$ and $\zz=(z_1,\ldots,z_v)$
to stand for the variables appearing only in $\psi$, where $d+u+v=n$.
\end{problem}

Since linear inequalities are trivially concave quadratic
polynomials, our algorithm (Algorithm $\igfqc$ in Section \ref{sec:alg}) can deal
with the linear case too. In fact, it is a generalization of the
algorithm for linear inequalities.

\oomit{Actually, since our main result (Theorem \ref{the:main}) is a generalization of Motzkin's theorem, our algorithm is essentially the same as other interpolant generation algorithms (e.g., \cite{RS10}) based on Motzkin's theorem when all the $f_i$ and $g_j$ are linear.

\begin{example}
For the formulas in Example \ref{exam:pre}, the input and output of our algorithm are
\begin{description} 
\item[{\sf In:}]   $\phi :f_1 = x_1, f_2 = x_2,g_1= -x_1^2+2 x_1 - x_2^2 + 2 x_2 - y^2$;\\
$\psi : f_3= -x_1^2-x_2^2 -2x_2-z^2$
\item[{\sf Out:}]  $\frac{1}{2}x_1^2+\frac{1}{2}x_2^2+2x_2 > 0$
\end{description}
\end{example}

In the following sections, we come up with an condition $\sosc$ (Definition \ref{def:sosc}) and generalize the
Motzkin's transposition theorem (Theorem \ref{motzkin-theorem}) to Theorem \ref{the:main} for concave quadratic case when this condition hold in Section \ref{theory}.
When the condition $\sosc$ holds, we give a method to solve Problem 1  based on Theorem \ref{the:main} in Section \ref{sec:hold}. For the general case of concave quadratic ,
 Problem 1 is solved in a recursive way in Section \ref{g:int}. }

The proposed algorithm is recursive: the base case is when no sum
of squares (SOS)
polynomial can be generated by a nonpositive constant combination of
nonstrict inequalities in $\phi \land \psi$.
When this condition is not satisfied, i.e., an SOS polynomial
can be generated by a nonpositive constant combination of
nonstrict inequalities in $\phi \land \psi$, then it is possible
to identify variables which can be eliminated by replacing them
by linear expressions in terms of other variables and thus
generate equisatisfiable problem with fewer variables on which
the algorithm can be recursively invoked.

\begin{lemma} \label{lemma:1}
Let $U \in \RR^{(n+1) \times (n+1)}$ be a matrix.
  If
  $\left<U,\left( \begin{matrix}
            1 & \xx^T \\
            \xx & \XX
          \end{matrix}
        \right ) \right >  \le 0$
  for any $\xx \in \RR^n$ and symmetric matrix $\XX \in \RR^{n\times n}$ with
  $\left( \begin{matrix}
            1 & \xx^T \\
            \xx & \XX
          \end{matrix} \right ) \succeq 0$
  , then $U \preceq 0$.
\end{lemma}
\begin{proof}
  Assume that $U \not \preceq 0$.
  Then there exists a column vector $\yy=(y_0,y_1,\ldots,y_n)^T \in \RR^{n+1}$ such that $c:=\yy^T U\yy=\left< U, \yy \yy^T \right>>0$. Denote $M= \yy \yy^T$, then $M \succeq 0$.

  If $y_0 \neq 0$, then let $\xx = (\frac{y_1}{y_0},\ldots,\frac{y_n}{y_0})^T$, and $\XX= \xx \xx^T$.
  Thus,
  $\left( \begin{matrix}
            1 & \xx^T \\
            \xx & \XX
          \end{matrix}
        \right ) =
        \left( \begin{matrix}
            1 & \xx^T \\
            \xx & \xx \xx^T
          \end{matrix}
        \right ) = \frac{1}{y_0^2} M \succeq $,
  and
  $\left<U,\left( \begin{matrix}
            1 & \xx^T \\
            \xx & \XX
          \end{matrix}
        \right ) \right> = \left<U, \frac{1}{y_0^2} M \right> = \frac{c}{y_0^2} >0$,
    which contradicts with
   $\left<U,\left( \begin{matrix}
            1 & \xx^T \\
            \xx & \XX
          \end{matrix}
        \right ) \right >  \le 0$.

  If $\yy_0= 0$, then $M_{(1,1)} = 0$.
  Let $M{'}=\frac{|U_{(1,1)}|+1 }{c} M$, then $M{'} \succeq 0$. Further,
  let $M{''} = M{'}+\left( \begin{matrix}
            1 & 0& \cdots & 0 \\
            0 & 0& \cdots & 0 \\
            \vdots & \vdots &\ddots &\vdots \\
            0 & 0& \cdots & 0
          \end{matrix}
        \right )$.
  Then $M{''}\succeq 0$ and $M{''}_{(1,1)}=1$.
  Let
  $\left( \begin{matrix}
            1 & \xx^T \\
            \xx & \XX
          \end{matrix}
        \right )= M{''} $, then
  \begin{align*}
    \left<U,\left( \begin{matrix}
            1 & \xx^T \\
            \xx & \XX
          \end{matrix}
    \right ) \right >&=
    \left<U, M{''} \right > =
    \left<U,
    M{'}+\left( \begin{matrix}
            1 & 0& \cdots & 0 \\
            0 & 0& \cdots & 0 \\
            \vdots & \vdots &\ddots &\vdots \\
            0 & 0& \cdots & 0
          \end{matrix}
    \right )
    \right >\\
    &=
    \left<U, \frac{|U_{(1,1)}|+1 }{c} M
    +\left( \begin{matrix}
            1 & 0& \cdots & 0 \\
            0 & 0& \cdots & 0 \\
            \vdots & \vdots &\ddots &\vdots \\
            0 & 0& \cdots & 0
          \end{matrix}
    \right )
    \right >  \\
    &=
    \frac{|U_{(1,1)}|+1 }{c}\left<U,  M
    \right > + U_{(1,1)}\\
    &=|U_{(1,1)}|+1+ U_{(1,1)} >0,
  \end{align*} which also contradicts with
   $\left<U,\left( \begin{matrix}
            1 & \xx^T \\
            \xx & \XX
          \end{matrix}
        \right ) \right >  \le 0$.
  Thus,
  the assumption does not hold, that is  $U\preceq 0$. \qed
\end{proof}

\begin{lemma} \label{lemma:2}
  Let $\mathcal{A} = \{ \yy \in \RR^m \mid A_i \yy-\va_i \ge 0, B_j \yy-\vb_j > 0,
  ~for~ i=1,\ldots,r, j=1,\ldots,
  \}$ be a nonempty set
  and $\mathcal{B} \subseteq \RR^m$ be an nonempty convex closed set. If
  $\mathcal{A} \cap \mathcal{B} = \emptyset$ and there does not exist a linear form $L(\yy)$
  such that
  \begin{align}  \label{separ:1}
    \forall \yy \in \mathcal{A}, L(\yy) >0, ~and~~
    \forall \yy \in \mathcal{B}, L(\yy) \le 0,
  \end{align}
  then there is a linear form $L_0(\yy) \not\equiv 0$
  and $\delta_1, \ldots, \delta_r \ge 0$
  such that
   \begin{align}
   L_0(\yy) = \sum_{i=1}^{r} \delta_i (A_i \yy - \alpha_i)~and~~
    \forall \yy \in \mathcal{B}, L_0(\yy) \le0.
  \end{align}
\end{lemma}
\begin{proof}
  Since $\mathcal{A}$ is defined by a set of linear inequalities, $\mathcal{A}$ is a
  convex set.
Using the separation theorem on disjoint convex sets,
  cf. e.g. \cite{barvinok02},
there exists a linear
  form $L_0(\yy) \not\equiv 0$ such that
  \begin{align}
    \forall \yy \in \mathcal{A}, L_0(\yy) \ge 0, ~and~~
    \forall \yy \in \mathcal{B}, L_0(\yy) \le0.
  \end{align}
  From (\ref{separ:1}) we have that
  \begin{align} \label{y0}
  \exists\yy_0 \in \mathcal{A}, ~~ L_0(\yy_0)=0.
  \end{align}
  Since
  \begin{align}
    \forall \yy \in \mathcal{A}, L_0(\yy) \ge 0,
  \end{align}
  then
  \begin{align*}
    &A_1 \yy -\alpha_1 \ge0\wedge \ldots \wedge A_r \yy -\alpha_r \ge0 \wedge \\
    &B_1 \yy -\beta_1 > 0\wedge \ldots \wedge B_s \yy -\beta_s > 0 \wedge -L_0(\yy) >0
  \end{align*}
  has no solution w.r.t. $\yy$.
  Using Corollary \ref{cor:linear}, there exist $\lambda_1,\ldots,\lambda_r \ge0$, $\eta_0, \ldots, \eta_s \ge0$ and $\eta \ge0$ such that
  \begin{align}
    &\sum_{i=1}^{r} \lambda_i (A_i\yy - \alpha_i) + \sum_{j=1}^{s} \eta_j (B_j \yy -\beta_j)+
    \eta (-L_0(\yy)) + \eta_0 \equiv 0, \label{eq:lem2} \\
    &\sum_{j=0}^{s} \eta_j + \eta > 0. \label{ineq:lem2}
  \end{align}
  Applying $\yy_0$ in (\ref{y0}) to (\ref{eq:lem2}) and (\ref{ineq:lem2}), it follows
  \begin{align*}
    \eta_0=\eta_1=\ldots=\eta_s=0,~~ \eta >0.
  \end{align*}
  For $i=1,\ldots,r$, let $\delta_i=\frac{\lambda_i}{\eta} \ge 0$, then
  \begin{align*}
   L_0(\yy) = \sum_{i=1}^{r} \delta_i (A_i \yy -\alpha_i)~and~~
    \forall \yy \in \mathcal{B}, L_0(\yy) \le0. ~~~ \qed
  \end{align*}
\end{proof}

The lemma below asserts the existence of a strict linear inequality
separating $\mathcal{A}$ and $\mathcal{B}$ defined above,
for the case when any nonnegative constant combination of the linearization of $f_i$s is
positive.

\begin{lemma} \label{linear}
Let $\mathcal{A} = \{ \yy \in \RR^m \mid A_i \yy-\va_i \ge 0, B_j \yy-\vb_j > 0,
  ~for~ i=1,\ldots,r, j=1,\ldots,
  \}$ be a nonempty set
  and $\mathcal{B} \subseteq \RR^m$ be an nonempty convex closed set, $\mathcal{A} \cap \mathcal{B} = \emptyset$.
  There exists a linear form $L(\xx,\overline{\XX})$ such that
  \begin{align*}
    \forall (\xx,\overline{\XX}) \in \mathcal{A}, L(\xx,\overline{\XX}) >0, ~and~~
    \forall (\xx,\overline{\XX}) \in \mathcal{B}, L(\xx,\overline{\XX}) \le0,
  \end{align*}
whenever there does not exist $\lambda_i \ge 0$, s.t.,
$ \sum_{i=1}^{r} \lambda_i P_i \preceq 0$.

  \end{lemma}

\begin{proof}
Proof is by contradiction.   \oomit{ i.e., there does not exists a linear form $L(\xx,\overline{\XX})$ such that
  \begin{align*}
    \forall (\xx,\overline{\XX}) \in \mathcal{A}, L(\xx,\overline{\XX}) >0, ~and~~
    \forall (\xx,\overline{\XX}) \in \mathcal{B}, L(\xx,\overline{\XX}) \le0.
  \end{align*}}
Given that $\mathcal{A}$ is defined by a set of linear inequalities
and $\mathcal{B}$ is a closed convex nonempty set,
  by Lemma \ref{lemma:2},
  there exist a linear form $L_0(\xx,\overline{\XX}) \not\equiv 0$
  and $\delta_1, \ldots, \delta_r \ge 0$
  such that
   \begin{align*}
   L_0(\xx,\overline{\XX}) = \sum_{i=1}^{r} \delta_i
   \left<P_i,\left( \begin{matrix}
            1 & \xx^T \\
            \xx & \XX
          \end{matrix}
        \right ) \right >~and~~
    \forall (\xx,\overline{\XX}) \in \mathcal{B}, L_0(\xx,\overline{\XX}) \le0.
  \end{align*}
  I.e. there exists an symmetrical matrix $\mathbf{L} \not\equiv 0$ such that
 \begin{align}
   &\left<\mathbf{L},\left( \begin{matrix}
            1 & \xx^T \\
            \xx & \XX
          \end{matrix}
        \right ) \right >
    \equiv \sum_{i=1}^{r} \delta_i
   \left<P_i,\left( \begin{matrix}
            1 & \xx^T \\
            \xx & \XX
          \end{matrix}
        \right ) \right >, \label{eq:mat}\\
   & \forall (\xx,\overline{\XX}) \in \mathcal{B},
    \left<\mathbf{L},\left( \begin{matrix}
            1 & \xx^T \\
            \xx & \XX
          \end{matrix}
        \right ) \right > \le0. \label{ineq:mat}
  \end{align}
  Applying Lemma \ref{lemma:1} to (\ref{ineq:mat}), it follows $\mathbf{L}
  \preceq 0$.  This implies that $\sum_{i=1}^{r} \delta_i P_i =\mathbf{L} \preceq 0$,
  which is in contradiction to
the assumption that there does not exist $\lambda_i \ge 0$, s.t.,
$ \sum_{i=1}^{r} \lambda_i P_i \preceq 0$
\qed
\end{proof}

\begin{definition} \label{def:sosc}
For given formulas $\phi$ and $\psi$ as in Problem 1,
it satisfies the  non-existence of an SOS condition ($\sosc$) iff
there do not exist $\delta_1\ge0, \ldots, \delta_r\ge 0$, such that
  $-(\delta_1 f_1 + \ldots + \delta_r f_r)$  is a non-zero SOS.
\end{definition}

\oomit{The above condition implies that there is no \textcolor{blue}{nonnegative} constant combination
of nonstrict inequalities which is always {nonpositive}.
\textcolor{green}{If quadratic polynomials appearing in $\phi$ and $\psi$ are
linearized, then the above condition is equivalent to requiring
that every nonnegative linear combination of the linearization of
nonstrict inequalities in $\phi$ and $\psi$ is {negative.}}
The following theorem gives a method for generating an
interpolant for this case by considering linearization of the
problem and using Corollary \ref{cor:matrix}. In that sense, this
theorem is a generalization of Motzkin's theorem to CQ polynomial inequalities. }

The following theorem gives a method for generating an
interpolant when the condition $\sosc$\ holds by considering linearization of the
problem and using Corollary \ref{cor:matrix}. In that sense, this
theorem is a generalization of Motzkin's theorem to CQ polynomial inequalities.

The following separation lemma about a nonempty convex set $\mathcal{A}$ generated by
linear inequalities that is disjoint from another nonempty closed
convex set $\mathcal{B}$ states that if there is no strict linear
inequality that holds over $\mathcal{A}$ and does not hold on any
element in $\mathcal{B}$, then there is a hyperplane separating
$\mathcal{A}$ and $\mathcal{B}$, which is a nonnegative linear
combination of nonstrict inequalities.

\begin{theorem}
  \label{the:main}
  Let $f_1,\ldots,f_r,g_1,\ldots,g_s$ are CQ polynomials
   and the $K$ is defined as in (\ref{eq:opt}) with $K=\emptyset$. If the condition $\sosc$ holds,
     then there exist $\lambda_i\ge 0$ ($i=1,\cdots,r$), $\eta_j \ge 0$ ($j=0,1,\cdots,s$) and a quadratic SOS polynomial $h \in \RR[\xx]$ such that
  \begin{align}
    &\sum_{i=1}^{r} \lambda_i f_i +\sum_{j=1}^{s} \eta_j g_j + \eta_0 + h \equiv 0,\\
    &\eta_0+\eta_1 + \ldots + \eta_s = 1.
  \end{align}
\end{theorem}

The proof uses the fact that if $f_i$s satisfy the $\sosc$
condition, then the linearization of $f_i$s and $g_j$s can be
exploited to generate an interpolant expressed in terms of $\xx$.
The main issue is to decompose the result from the
linearized problem into two components giving an interpolant.

\begin{proof}
Recall from Section \ref{linearization} that
  \begin{align*}
    f_i = \left<P_i,\left( \begin{matrix}
            1 & \xx^T \\
            \xx & \xx\xx^T
          \end{matrix}
        \right ) \right > ,~~
    g_j = \left<Q_j,\left( \begin{matrix}
            1 & \xx^T \\
            \xx & \xx\xx^T
          \end{matrix}
        \right ) \right >.
  \end{align*}
  Let
  \begin{equation}\label{eq:mom}
    \begin{aligned}
    &\mathcal{A}:=\{ (\xx,\overline{\XX}) \mid
    \wedge_{i=1}^r \left<P_i,\left( \begin{matrix}
            1 & \xx^T \\
            \xx & \XX
          \end{matrix}
        \right ) \right > \ge 0,
    \wedge_{j=1}^s \left<Q_j,\left( \begin{matrix}
            1 & \xx^T \\
            \xx & \XX
          \end{matrix}
        \right ) \right > > 0
        \oomit{~for~
        \begin{matrix}
            i=1,\ldots,r \\
            j=1,\ldots,s
          \end{matrix} }
    \}, \\
      &\mathcal{B}:=\{(\xx,\overline{\XX})\mid   \left( \begin{matrix}
          1 & \xx^T\\
          \xx & \XX
        \end{matrix}
      \right)\succeq 0 \},  \\
    \end{aligned}
  \end{equation}
  be linearizations of the CQ polynomials  $f_i$s
and $g_j$s,
  where  \begin{align*}
  \overline{\XX}=(&\XX_{(1,1)},\XX_{(2,1)},\XX_{(2,2)},\ldots,
    \XX_{(k,1)}, \ldots,\XX_{(k,k)}, \ldots, \XX_{(n,1)}, \ldots,\XX_{(n,n)}).
  \end{align*}

By Theorem \ref{the:1}, $\mathcal{A} \cap \mathcal{B} =K_1=K=\emptyset$.

Since $f_i$s satisfy the $\sosc$ condition, its linearization
satisfy the condition of Lemma \ref{linear}; thus
there exists  a linear form $\mathcal{L}(\xx,\XX)=\left<L,\left( \begin{matrix}
        1 & \xx^T \\
        \xx & \XX
      \end{matrix}
    \right )  \right>$ such that
  \begin{align}
    & \mathcal{L}(\xx,\XX) > 0, ~for~ (\xx,\XX) \in \mathcal{A}, \label{lin-sep1}\\
    & \mathcal{L}(\xx,\XX) \le 0, ~for~ (\xx,\XX) \in \mathcal{B} \label{lin-sep2}.
  \end{align}

Applying Lemma \ref{lemma:1}, it follows $L \preceq 0$.
  Additionally, applying Lemma \ref{cor:matrix} to
  (\ref{lin-sep1}) and denoting $-L$ by $P_0$,
  there exist $\overline{\lambda_0}, \overline{\lambda_1},\ldots,\overline{\lambda_r} \ge 0$
and $\overline{\eta_0},\overline{\eta_1},\ldots,\overline{\eta_s} \ge 0$ such that
\begin{align*}
  & \sum_{i=0}^{r}\overline{\lambda_i} \left<P_i,\left( \begin{matrix}
            1 & \xx^T \\
            \xx & \XX
          \end{matrix}
        \right ) \right > +
  \sum_{j=1}^{s}\overline{\eta_j} \left<Q_j,\left( \begin{matrix}
            1 & \xx^T \\
            \xx & \XX
          \end{matrix}
        \right ) \right > + \overline{\eta_0} \equiv 0,  \\
  & \overline{\eta_0} + \overline{\eta_1} + \ldots + \overline{\eta_s} > 0.
\end{align*}
Let $\lambda_i=\frac{\overline{\lambda_i}}{\sum_{j=0}^s \overline{\eta_j}}$,
$\eta_j=\frac{\overline{\eta_j}}{\sum_{j=0}^s \overline{\eta_j}}$,  then
\begin{align}
  & \lambda_0
        \left<-U,\left( \begin{matrix}
            1 & \xx^T \\
            \xx & \XX
          \end{matrix}
        \right ) \right >+
  \sum_{i=1}^{r}\lambda_i \left<P_i,\left( \begin{matrix}
            1 & \xx^T \\
            \xx & \XX
          \end{matrix}
        \right ) \right > +
  \sum_{j=1}^{s}\eta_j \left<Q_j,\left( \begin{matrix}
            1 & \xx^T \\
            \xx & \XX
          \end{matrix}
        \right ) \right > + \eta_0 \equiv 0, \label{lin-sep5}\\
  & \eta_0 + \eta_1 + \ldots + \eta_s =1. \label{lin-sep6}
\end{align}
Since for any $\xx$ and symmetric matrix $\XX$, (\ref{lin-sep5}) holds,  by setting $\XX=\xx\xx^T$,
\begin{align*}
\lambda_0
        \left<-U,\left( \begin{matrix}
            1 & \xx^T \\
            \xx & \xx\xx^T
          \end{matrix}
        \right ) \right >+
   \sum_{i=1}^{r}\lambda_i \left<P_i,\left( \begin{matrix}
            1 & \xx^T \\
            \xx & \xx\xx^T
          \end{matrix}
        \right ) \right > +
  \sum_{j=1}^{s}\eta_j \left<Q_j,\left( \begin{matrix}
            1 & \xx^T \\
            \xx & \xx\xx^T
          \end{matrix}
        \right ) \right > + \eta_0 \equiv 0,
\end{align*}
which means that
\begin{align*}
     h+\sum_{i=1}^{r} \lambda_i f_i +\sum_{j=1}^{s} \eta_j g_j + \eta_0  \equiv 0,
  \end{align*}
where $h=\lambda_0 \left<-U,\left( \begin{matrix}
            1 & \xx^T \\
            \xx & \xx\xx^T
          \end{matrix}
        \right ) \right >$. Since $U\preceq 0$,  $-U \succeq 0$. Hence $h$ is a quadratic SOS polynomial.
\qed
\end{proof}

\subsection{Base Case: Generating Interpolant when NSOSC is satisfied} \label{sec:hold}
Using the above theorem, it is possible to generate an
interpolant for $\phi$  and $\psi$  from the SOS polynomial $h$ obtained using
the theorem which can be split into two SOS polynomials in the
common variables of $\phi$ and $\psi$.
This is
proved in the following theorem using some lemma as follows.

\begin{lemma} \label{h:sep}
Given a quadratic SOS polynomial
  $h(\xx,\yy,\zz) \in \RR[\xx,\yy,\zz]$ on variables
  $\xx=(x_1,\cdots,x_d) \in
  \RR^{d}$,$\yy=(y_1,\cdots,y_u)
  \in \RR^{u}$
  and $\zz=(z_1,\cdots,z_v) \in \RR^{v}$ such that the coefficients of
  $y_i z_j$ ($i=1,\cdots,u,j=1,\cdots,v$) are all vanished when
  expanding $h(\xx,\yy,\zz)$, there exist two quadratic
  polynomial $h_1(\xx,\yy) \in \RR[\xx,\yy]$ and
  $h_2(\xx,\zz) \in \RR[\xx,\zz]$ such that $h=h_1+h_2$,
  moreover, $h_1$ and $h_2$ both are SOS.
\end{lemma}

\begin{proof}
  Since $h(\xx,\yy,\zz)$ is a quadratic polynomial and the coefficients of $y_i z_j$ ($i=1,\cdots,u,j=1,\cdots,v$) are all vanished when expanding $h(\xx,\yy,\zz)$, we have
  \begin{align*}
    h(\xx,\yy_1,\cdots,\yy_u,\zz)=a_1 y_1^2 + b_1(\xx,y_2,\cdots,y_u) y_1 + c_1(\xx,y_2,\cdots,y_u,\zz),
  \end{align*}
  where $a_1 \in \RR$, $b_1(\xx,y_2,\cdots,y_u) \in \RR[\xx,y_2,\cdots,y_u]$ is a linear function and $c_1(\xx,y_2,\cdots,y_u,\zz)\in \RR[\xx,y_2,\cdots,y_u,\zz]$ is a
  quadratic polynomial.
  Since $h(\xx,\yy,\zz)$ is an SOS polynomial, so
  \begin{align*}
    \forall (\xx,y_1,\cdots,y_u,\zz) \in \RR^{d+u+v} ~~~ h(\xx,y_1,\cdots, y_u, \zz) \geq 0.
  \end{align*}
  Thus $a_1=0 \wedge b_1 \equiv 0$ or $a_1>0$.
  If $a_1=0 \wedge b_1 \equiv 0$ then we denote
  \begin{align*}
    p_1(\xx,y_2,\cdots,y_u,\zz)=c_1(\xx,y_2,\cdots,y_u,\zz), ~~ q_1(\xx,y_1,\cdots,y_u)=0;
  \end{align*}
  otherwise, $a_1 >0$, and  we denote
 {\small  \begin{align*}
    p_1(\xx,y_2,\cdots,y_u,\zz)=h(\xx,-\frac{b_1}{2 a_1},y_2,\cdots,y_u,\zz),~~
    q_1(\xx,y_1,\cdots,y_u)=a_1 (y_1  + \frac{b_1}{2 a_1})^2.
  \end{align*} }
  Then, it is easy to see $p_1(\xx,y_2,\cdots,y_u,\zz)$ is a quadratic polynomial satisfying
  \begin{align*}
    h(\xx,y_1,\cdots,y_u,\zz)= p_1(\xx,y_2,\cdots,y_u,\zz)+q_1(\xx,y_1,\cdots,y_u),
  \end{align*}
  and
  \begin{align*}
    \forall (\xx,y_2,\cdots,y_u,\zz) \in \RR^{r+s-1+t} ~~~ p_1(\xx,y_2,\cdots, y_u, \zz) \geq 0,
  \end{align*}
  moreover, the coefficients of $y_i z_j$ ($i=2,\cdots,s,j=1,\cdots,t$) are all vanished when expanding  $p_1(\xx,y_2,\cdots,y_u,\zz)$,  and $q_1(\xx,y_1,\cdots,y_u)\in \RR[\xx,\yy]$ is an SOS.
  With the same reason, we can obtain $p_2(\xx,y_3,\cdots,y_u,\zz)$, $\cdots$, $p_u(\xx,\zz)$ and
  $q_2(\xx,y_2, \cdots,y_u)$, $\cdots$, $q_s(\xx,y_u)$ such that
  \begin{align*}
    p_{i-1}(\xx,y_{i},\cdots,y_u,\zz) = p_{i}(\xx,y_{i+1},\cdots,y_u,\zz)+
    q_i(\xx,y_{i},\cdots,y_u), \\[2mm]
    \forall (\xx,y_{i+1},\cdots,y_u,\zz) \in \RR^{d+u-i+v} ~ p_i(\xx,y_{i+1},\cdots, y_u, \zz) \geq 0,\\
    q_i(\xx,y_{i},\cdots,y_u) {\rm~ is ~ a ~ SOS ~polynomial},
  \end{align*}
  for $i=2,\cdots,u$.
  Therefore, let
  \begin{align*}
    h_1(\xx,\yy)=q_1(\xx,y_1,\cdots,y_u) + \cdots + q_s(\xx,y_u), ~~
    h_2(\xx,\zz)=p_u(\xx,\zz),
  \end{align*}
  we have $h_1(\xx,\yy) \in \RR[\xx,\yy]$ is an SOS and
    $\forall (\xx,\zz) \in \RR^{r+t} ~ h_2(\xx,\zz)= p_u(\xx,\zz) \geq 0$.
  Hence,  $h_2(\xx,\zz)$ is also an SOS, because that for the case of degree $2$, a polynomial is
  positive semi-definite iff it is an SOS polynomial.
  Thus $h_1(\xx,\yy) \in \RR[\xx,\yy]$ and $h_2(\xx,\zz) \in \RR[\xx,\zz]$ are both SOS,
  moreover,
{\small   \begin{align*}
    h_1+h_2=q_1+\cdots +q_{u-1}+q_u +p_u
    =q_1+\cdots+ q_{u-1} +p_{u-1}
    =&\cdots
    = q_1+p_1
    =h. ~~ \qed
  \end{align*} }
\end{proof}

The above proof of Lemma \ref{h:sep} gives a method to express
$h, h_1, h_2$ as sums of squares of linear expressions and a
nonnegative real number.

\begin{lemma}\label{lem:split}
Let $h, h_1, h_2$ be as in the statement of Lemma \ref{h:sep}. Then,
{\small \begin{align*}
  \mathrm{(H)}:~ h(\xx,\yy,\zz)=&a_1 (y_1 -l_1(\xx,y_2,\ldots,y_u))^2 + \ldots +  a_u (y_u -l_u(\xx))^2+\\
    &a_{u+1} (z_1 -l_{u+1}(\xx,z_2,\ldots,z_v))^2 + \ldots +  a_{u+v} (z_v -l_{u+v}(\xx))^2+\\
    &a_{u+v+1}(x_1 - l_{u+v+1}(x_2,\ldots,x_d))^2 + \ldots + a_{u+v+d} (x_d - l_{u+v+d})^2 \\
    &+a_{u+v+d+1},
\end{align*} }
where $a_i \ge 0$ and $l_j$ is a linear expression
 in the corresponding  variables, for $ i=1,\ldots,u+v+d+1$, $
 j=1,\ldots,u+v+d$. Further,
{\small \begin{align*}
  \mathrm{(H1)}:~ &h_1(\xx,\yy)=a_1 (y_1 -l_1(\xx,y_2,\ldots,y_u))^2 + \ldots +  a_u (y_u -l_u(\xx))^2+\\
    &\frac{a_{u+v+1}}{2}(x_1 - l_{u+v+1}(x_2,\ldots,x_d))^2 + \ldots + \frac{a_{u+v+d}}{2} (x_d - l_{u+v+d})^2 +\frac{a_{u+v+d+1}}{2}, \\[2mm]
  \mathrm{(H2)}:~ &h_2(\xx,\zz)=a_{u+1} (z_1 -l_{u+1}(\xx,z_2,\ldots,z_v))^2 + \ldots +  a_{u+v} (\zz_v -l_{u+v}(\xx))^2+\\
    &\frac{a_{u+v+1}}{2}(x_1 - l_{u+v+1}(x_2,\ldots,x_d))^2 + \ldots + \frac{a_{u+v+d}}{2} (x_d - l_{u+v+d})^2+\frac{a_{u+v+d+1}}{2}.
\end{align*} }
\end{lemma}

\begin{theorem} \label{the:int}
   Let $\phi$ and $\psi$ as defined in Problem 1 with $\phi\wedge\psi\models\bot$,  which satisfy
   $\sosc$.
   Then there exist $\lambda_i\ge 0$ ($i=1,\cdots,r$), $\eta_j \ge 0$ ($j=0,1,\cdots,s$) and two quadratic SOS polynomial $h_1 \in \RR[\xx,\yy]$ and $h_2 \in \RR[\xx,\zz]$ such that
  \begin{align}
    & \sum_{i=1}^{r} \lambda_i f_i +\sum_{j=1}^{s} \eta_j g_j + \eta_0 + h_1+h_2 \equiv 0, \label{con1:inte}\\
    & \eta_0+\eta_1 + \ldots + \eta_s = 1.\label{con2:inte}
  \end{align}
  Moreover,  if $\sum_{j=0}^{s_1} \eta_j > 0$, then $I >0$ is an interpolant,
  otherwise $I \ge 0$ is an interpolant, where $I = \sum_{i=1}^{r_1}
  \lambda_i f_i +\sum_{j=1}^{s_1} \eta_j g_j + \eta_0 + h_1 \in \RR[\xx]$.
\end{theorem}
\begin{proof}
From Theorem \ref{the:main}, there exist $\lambda_i\ge 0$ ($i=1,\cdots,r$), $\eta_j \ge 0$ ($j=0,1,\cdots,s$) and a quadratic SOS polynomial $h \in \RR[\xx,\yy,\zz]$ such that
  \begin{align}
    & \sum_{i=1}^{r} \lambda_i f_i +\sum_{j=1}^{s} \eta_j g_j + \eta_0 + h \equiv 0, \label{r1}\\
    & \eta_0+\eta_1 + \ldots + \eta_s = 1. \label{r2}
  \end{align}
Obviously, (\ref{r1}) is equivalent to the following formula
\begin{align*}
 \sum_{i=1}^{r_1} \lambda_i f_i +\sum_{j=1}^{s_1} \eta_j g_j +\eta_0+
 \sum_{i=r_1+1}^{r} \lambda_i f_i +\sum_{j=s_1+1}^{s} \eta_j g_j+   h \equiv 0,
\end{align*}
It's easy to see that
\begin{align*}
 \sum_{i=1}^{r_1} \lambda_i f_i +\sum_{j=1}^{s_1} \eta_j g_j +\eta_0  \in \RR[\xx,\yy], ~~
 \sum_{i=r_1+1}^{r} \lambda_i f_i +\sum_{j=s_1+1}^{s} \eta_j g_j  \in \RR[\xx,\zz].
\end{align*}
Thus, for any $1\le i \le u$, $1\le j \le v$, the term $\yy_i \zz_j$ does not appear in
\begin{align*}
 \sum_{i=1}^{r_1} \lambda_i f_i +\sum_{j=1}^{s_1} \eta_j g_j +\eta_0+
 \sum_{i=r_1+1}^{r} \lambda_i f_i +\sum_{j=s_1+1}^{s} \eta_j g_j .
\end{align*}
Since all the conditions in Lemma \ref{h:sep} are satisfied, there exist two quadratic
SOS polynomial $h_1 \in \RR[\xx,\yy]$ and $h_2 \in \RR[\xx,\zz]$ such that
$h=h_1+h_2$.
Thus, we have
\begin{align*}
 &\sum_{i=1}^{r_1} \lambda_i f_i +\sum_{j=1}^{s_1} \eta_j g_j +\eta_0+h_1  \in \RR[\xx,\yy],\\
 &\sum_{i=r_1+1}^{r} \lambda_i f_i +\sum_{j=s_1+1}^{s} \eta_j g_j+h_2  \in \RR[\xx,\zz],\\
 &\sum_{i=1}^{r_1} \lambda_i f_i +\sum_{j=1}^{s_1} \eta_j g_j +\eta_0+h_1+
 \sum_{i=r_1+1}^{r} \lambda_i f_i +\sum_{j=s_1+1}^{s} \eta_j g_j+h_2 \equiv 0
\end{align*}
Besides, as
\begin{align*}
  I=\sum_{i=1}^{r_1} \lambda_i f_i +\sum_{j=1}^{s_1} \eta_j g_j +\eta_0+h_1 
=-(\sum_{i=r_1+1}^{r} \lambda_i f_i +\sum_{j=s_1+1}^{s} \eta_j g_j+h_2), 
\end{align*}
we have $I \in \RR[\xx]$.
It is easy to see that
\begin{itemize}
\item if  $\sum_{j=0}^{s_1} \eta_j > 0$ then
  $\phi \models I >0$  and $\psi \wedge I>0 \models \bot$,
    so $I > 0$ is an interpolation; and
  \item  if $\sum_{j=s_1+1}^{s} \eta_j > 0$ then
   $\phi \models I \ge 0$ and $\psi \wedge I\ge 0 \models \bot$,
 hence  $I \ge 0$ is an interpolation. \qed
\end{itemize}
\oomit{   Since $\sum_{j=0}^s \eta_j =1$ implies either  $ \sum_{j=0}^{s_1} \eta_j > 0$
   or $\sum_{j=s_1+1}^{s} \eta_j > 0$.
\qed}
\end{proof}

\subsection{Computing Interpolant using Semi-Definite Programming} \label{sec:sdp}



\oomit{When the condition $\sosc$ hold, from Theorem \ref{the:int} we can see that, the problem of
interpolant generation can be reduced to the following constraint solving problem.

{\em Find:
\begin{center}
real numbers $\lambda_i\ge 0~(i=1,\cdots,r)$, $\eta_j \ge 0~(j=0,1,\cdots,s)$, and\\
two quadratic SOS polynomials $h_1 \in \RR[\xx,\yy]$ and $h_2 \in \RR[\xx,\zz]$,
\end{center}
such that
  \begin{align*}
    & \sum_{i=1}^{r} \lambda_i f_i +\sum_{j=1}^{s} \eta_j g_j + \eta_0 + h_1+h_2 \equiv 0, \\
    & \eta_0+\eta_1 + \ldots + \eta_s = 1.
  \end{align*} } }

Below, we formulate computing $\lambda_i$s, $\eta_j$s and $h_1$
and $h_2$ as a semi-definite programming problem.

Let
\[W=\left( \begin{matrix}
           1 & \xx^T & \yy^T & \zz^T\\
            \xx & \xx\xx^T & \xx\yy^T & \xx\zz^T\\
            \yy & \yy\xx^T & \yy\yy^T & \yy\zz^T\\
            \zz & \zz\xx^T & \zz\yy^T & \zz\zz^T
          \end{matrix}
        \right )\]
\begin{align} \label{pq:def}
    f_i = \langle P_i, W\rangle  ,~~ g_j = \langle Q_j, W\rangle,
\end{align}
where $P_i$ and $Q_j$ are $(1+d+u+v) \times (1+d+u+v)$ matrices,
and
\begin{align*}
    h_1 = \langle M, W\rangle , ~~     h_2 = \langle \hat{M}, W\rangle,
\end{align*}
where $M=(M_{ij})_{4\times4}, \hat{M}=(\hat{M}_{ij})_{4\times4}$
 with appropriate dimensions, for example $M_{12} \in \RR^{1 \times d}$ and
$\hat{M}_{34} \in \RR^{u \times v}$.
Then, with $\sosc$,  by Theorem~\ref{the:int}, Problem 1 is reduced to
 the following $\sdp$ feasibility problem.

\textbf{Find:}
\begin{center}
$\lambda_1,\ldots, \lambda_r,\eta_0,\ldots,\eta_s \in \RR$ and real symmetric matrices $M, \hat{M} \in \RR^{(1+d+u+v)\times (1+d+u+v)}$
\end{center}
subject to
\begin{eqnarray*}
  \left\{ ~ \begin{array}{l}
  \sum_{i=1}^{r} \lambda_i P_i +\sum_{j=1}^{s} \eta_j Q_j + \eta_0 E_{1,1} + M+\hat{M} = 0$, $\sum_{j=0}^{s} \eta_j=1,\\[1mm]
  M_{41}=(M_{14})^T=0,M_{42}=(M_{24})^T=0,M_{43}=(M_{34})^T=0,M_{44}=0,\\[1mm]
  \hat{M}_{31}=(\hat{M}_{13})^T=0,\hat{M}_{32}=(\hat{M}_{23})^T=0,\hat{M}_{33}=0,\hat{M}_{34}=(\hat{M}_{43})^T=0,\\[1mm]
  M\succeq 0, \hat{M}\succeq 0,\lambda_i \ge0, \eta_j \ge 0, \mbox{ for }
  i=1,\ldots,r,j=0,\ldots,s,
  \end{array}
  \right.
  \end{eqnarray*}
  where $E_{1,1}$ is a $(1+d+u+v) \times (1+d+u+v)$ matrix, whose $(1,1)$ entry is $1$ and the others are $0$.

  This is a standard $\sdp$ feasibility problem, which can be
  solved efficiently by well known $\sdp$ solvers, e.g., CSDP
  \cite{CSDP}, 
  SDPT3
  \cite{SDPT3}, SeDuMi \cite{SeDuMi}, etc., with
  time complexity polynomial in $n=d + u + v$.

\begin{remark} \label{remark:1}
Problem 1 is a typical quantifier elimination (QE) problem, which can be solved symbolically. However, it is very hard to solve large problems by general QE algorithms because of their high complexity. So, reducing Problem 1 to $\sdp$ problem makes it possible to solve many large problems in practice. Nevertheless, one may doubt whether we can use numerical result in verification. We think that verification must be rigorous and numerical results should be verified first. For example, after solving the above $\sdp$ problem numerically, we verify that whether $-(\sum_{i=1}^{r} \lambda_i f_i +\sum_{j=1}^{s} \eta_j g_j + \eta_0)$ is an SOS by the method of Lemma \ref{lem:split}, which is easy to do. If it is, the result is guaranteed and output. If not, the result is unknown (in fact, some other techniques can be employed in this case, which we do not discuss in this paper.). Thus, our algorithm is sound but not complete.
\end{remark}

\subsection{General Case}

The case of
$\textit{Var}(\phi) \subset \textit{Var}(\psi)$ is
not an issue since $\phi$ serves as an interpolant of $\phi$
and $\psi$.
We thus assume
that $\textit{Var}(\phi) \nsubseteq \textit{Var}(\psi)$.
We show below how an interpolant can be generated in the general case.
If $\phi$ and $\psi$ do not satisfy the $\sosc$
condition, i.e., an SOS polynomial $h(\xx, \yy, \zz)$ can be computed from nonstrict
inequalities $f_i$s using nonpositive constant multipliers, then
by the lemma below, we can construct ``simpler'' interpolation
subproblems $\phi', \psi'$
from $\phi$ and $\psi$ by constructing from $h$ an SOS polynomial
$f(\xx)$ such that $\phi \models f(\xx) \ge 0$ as well as $\psi
\models - f(\xx) \ge 0$.
Each $\phi'$ $\psi'$ pair has the following characteristics
because of which the algorithm
is recursively applied to $\phi'$ and $\psi'$.
\begin{enumerate}
\item[(i)] $\phi' \wedge \psi' \models \bot$,
\item[(ii)]  $\phi',\psi'$ have the same form as $\phi,\psi$, i.e., $\phi'$ and $\psi'$
are defined by some $f_i' \ge 0$ and $g_j'>0$, where $f_i'$ and $g_j'$ are CQ,
\item[(iii)] $\#(\textit{Var}(\phi') \cup \textit{Var}(\psi')) <
  \#(\textit{Var}(\phi) \cup \textit{Var}(\psi))$ to ensure
  termination of the recursive algorithm, and
\item[(iv)] an interpolant
$I$ for $\phi$ and $\psi$ can be computed
from an interpolant $I'$ for $\phi'$ and $\psi'$ using $f$.
\end{enumerate}

\oomit{Now, suppose $\textit{Var}(\phi) \nsubseteq \textit{Var}(\psi)$ and the condition $\sosc$ does not hold,
i.e., there exist
$\delta_1,\ldots,\delta_r \ge 0$ such that
$(-\sum_{i=1}^r \delta_i f_i)$ is a nonzero SOS polynomial, we construct such
$\phi'$ and $\psi'$ satisfy $(i)-(iv)$ below. }

\begin{lemma} \label{lemma:dec}
  If Problem 1 does not satisfy the $\sosc$ condition, there exists $f \in \RR[\xx]$,
  such that
  $\phi \Leftrightarrow \phi_1 \vee \phi_2$ and
  $\psi \Leftrightarrow \psi_1 \vee \psi_2$,
where,
\begin{align}
  &\phi_1= (f > 0 \wedge \phi) ,~~
  \phi_2 = (f = 0 \wedge \phi),\label{phi2}\\
  &\psi_1 = (-f > 0 \wedge \psi),~~
  \psi_2 =  (f = 0 \wedge \psi).\label{psi2}
\end{align}
\end{lemma}
\begin{proof}
Since $\sosc$ does not hold, there exist $\delta_1,\ldots, \delta_r \in \RR^+$ such that $-\sum_{i=1}^r \delta_i f_i$ is a nonzero SOS. Let $h(\xx,\yy,\zz)$ denote this quadratic SOS polynomial.

Since $(-\sum_{i=1}^{r_1} \delta_i f_i) \in \RR[\xx,\yy]$
and $(-\sum_{i=r_1+1}^r \delta_i f_i) \in \RR[\xx,\zz]$, the coefficient of any term
$\yy_i \zz_j, 1\le i \le u,1 \le j \le v,$ is 0 after expanding
$h$. By Lemma \ref{h:sep}
there exist two quadratic SOS polynomials $h_1 \in \RR[\xx,\yy]$ and
$h_2 \in \RR[\xx,\zz]$ such that $h=h_1+h_2$
with the following form:
{\small  \begin{align*}
  \mathrm{(H1)}: & ~ h_1(\xx,\yy)=a_1 (\yy_1 -l_1(\xx,\yy_2,\ldots,\yy_u))^2 + \ldots +  a_u (\yy_u -l_u(\xx))^2+\\
    &\frac{a_{u+v+1}}{2}(\xx_1 - l_{u+v+1}(\xx_2,\ldots,\xx_d))^2 + \ldots + \frac{a_{u+v+d}}{2} (\xx_d - l_{u+v+d})^2 +\frac{a_{u+v+d+1}}{2}, \\[3mm]
  \mathrm{(H2)}:& ~ h_2(\xx,\zz)=a_{u+1} (\zz_1 -l_{u+1}(\xx,\zz_2,\ldots,\zz_v))^2 + \ldots +  a_{u+v} (\zz_v -l_{u+v}(\xx))^2+\\
    &\frac{a_{u+v+1}}{2}(\xx_1 - l_{u+v+1}(\xx_2,\ldots,\xx_d))^2 + \ldots + \frac{a_{u+v+d}}{2} (\xx_d - l_{u+v+d})^2+\frac{a_{u+v+d+1}}{2}.
\end{align*} }
Let
\begin{align} \label{f:form}
f= \sum_{i=1}^{r_1} \delta_i f_i + h_1=-\sum_{i=r_1+1}^r \delta_i f_i-h_2.
\end{align}
Obviously, $f \in \RR[\xx,\yy]$ and $f \in \RR[\xx,\zz]$,
 this implies $f \in \RR[\xx]$.

Since $h_1, h_2$ are SOS, it is easy to see that
  $\phi \models f(\xx) \ge 0, ~~ \psi \models -f(\xx) \ge 0$.
Thus,
  $\phi \Leftrightarrow \phi_1 \vee \phi_2$,
  $\psi \Leftrightarrow \psi_1 \vee \psi_2$.
\qed
\end{proof}

Using the above lemma, an interpolant $I$ for $\phi$ and $\psi$
can be constructed from an interpolant $I_{2,2}$ for $\phi_2$ and
$\psi_2$.

\begin{theorem} \label{lemma:p22}
Let $\phi$, $\psi$, $\phi_1, \phi_2, \psi_1, \psi_2$ as defined in
  Lemma \ref{lemma:dec}, then  given an interpolant  $I_{2,2}$ for $\phi_2$ and $\psi_2$,
    $I:= (f >0 ) \vee (f \ge0 \wedge I_{2,2})$
  is an interpolant for $\phi$ and $\psi$.
\end{theorem}

\begin{proof}
  It is easy to see that $f > 0$ is an interpolant for both
$(\phi_1, \psi_1)$ and $(\phi_1, \psi_2)$, and
$f \ge 0$ is an interpolant for $(\phi_2, \psi_1)$.
Thus, if $I_{2,2}$ is an interpolant for $(\phi_2,\psi_2)$, then $I$
  is an interpolant for $\phi$ and $\psi$. \qed
\end{proof}

An interpolant for $\phi_2$ and $\psi_2$ is constructed
recursively since the new constraint included in $\phi_2$
(similarly, as well
as in $\psi_2$) is: $\sum_{i=1}^{r_1} \delta_i f_i + h_1=0$ with
$h_1$ being an SOS.
Let $\phi'$ and $\psi'$ stand for the formulas constructed
after analyzing $\phi_2$ and $\psi_2$ respectively.
Given that $\delta_i$ as well as $f_i \ge
0$ for each $i$, case analysis is performed on $h_1$ depending upon
whether it has a positive constant
$a_{u+v+d+1} > 0$  or not.

\begin{theorem} \label{the:gcase:1}
Let   $\phi'\hat{=} (0>0)$ and $\psi'\hat{=} (0>0)$.  In the proof of Lemma \ref{lemma:dec},  if $a_{u+v+d+1} > 0$, then
  $\phi'$ and $\psi'$ satisfy  $(i)-(iv)$.
\end{theorem}
\begin{proof}
   $(i),(ii)$ and $(iii)$ are trivially satisfied.
  Since $a_{u+v+d+1} > 0$, it is easy to see that
  $h_1 >0$ and $h_2 >0$.
  From (\ref{phi2}), (\ref{psi2}) and (\ref{f:form}), we have
  $\phi_2 \models h_1 = 0$, and $\psi_2 \models h_2=0$.
  Thus $\phi_2 \Leftrightarrow \phi' \Leftrightarrow\bot$ and
  $\psi_2 \Leftrightarrow \psi' \Leftrightarrow\bot$.
\qed
\end{proof}

\oomit{For the case $a_{u+v+d+1} > 0$, we construct $\phi'$ and $\psi'$ in Theorem \ref{the:gcase:1},
then we construct $\phi'$ and $\psi'$ on the case $a_{u+v+d+1} = 0$ below.}

In case 
 $a_{u+v+d+1} = 0$, from the fact that $h_1$ is an SOS and has
 the form $\mathrm{(H1)}$, each nonzero square term in $h_1$ is
 identically 0. This implies that some of the variables in $\xx, \yy$ can be
linearly expressed in term of other variables; the same argument applies
to $h_2$ as well.  In particular, at least
 one variable is eliminated in both $\phi_2$ and $\psi_2$,
 reducing the number of variables appearing in $\phi$ and
 $\psi$, which ensures the termination of the algorithm. A
 detailed analysis is given in following lemmas, where it is shown how
 this elimination of variables is performed,
generating $\phi'$ and $\psi'$ on which the algorithm can be
recursively invoked; an a theorem is also proved to ensures this.

\begin{lemma} \label{lemma:elim}
In the proof of  Lemma \ref{lemma:dec}, if $a_{u+v+d+1} = 0$, then  $\xx$ can be
represented as $(\xx^1, \xx^2)$, $\yy$ as $(\yy^1, \yy^2)$ and $\zz$ as $(\zz^1, \zz^2)$,
 such that
\begin{align*}
  &\phi_2 \models ( (\yy^1 = \Lambda_1 \left( \begin{matrix} \xx^2 \\ \yy^2 \end{matrix} \right) + \gamma_1)\wedge (\xx^1 = \Lambda_3 \xx^2 + \gamma_3) ), \\
  &\psi_2 \models  ((\zz^1 = \Lambda_2 \left( \begin{matrix} \xx^2 \\ \zz^2 \end{matrix} \right) + \gamma_2)\wedge (\xx^1 = \Lambda_3 \xx^2 + \gamma_3) ),
\end{align*}
and $\#(\textit{Var}(\xx^1)+\textit{Var}(\yy^1)+\textit{Var}(\zz^1)) > 0$,
 for matrixes $\Lambda_1,\Lambda_2,\Lambda_3$ and vectors $\gamma_1,\gamma_2,\gamma_3$.
\end{lemma}
\begin{proof}
From (\ref{phi2}), (\ref{psi2}) and (\ref{f:form}) we have
\begin{align} \label{phi-psi:2}
  \phi_2 \models h_1 = 0, ~~~~ \psi_2 \models h_2=0.
\end{align}
Since $h_1+h_2 =h$ is a nonzero polynomial, $a_{u+v+d+1} = 0$ ,
then there exist some $a_i \neq 0$, i.e. $a_i > 0$, for $1\le i \le u+v+d$.
Let
\begin{align*}
  &N_1:=\{i \mid a_i > 0 \wedge 1 \le i \le u \}, \\
  &N_2:=\{i \mid a_{u+i} > 0 \wedge 1 \le i \le v \}, \\
  &N_3:=\{i \mid a_{u+v+i} > 0 \wedge 1 \le i \le d \}.
\end{align*}
Thus,  $N_1$, $N_2$ and $N_3$ cannot all be empty. In addition,  $h_1=0 $ implies that
\begin{align*}
  &\yy_i=l_i(\xx,\yy_{i+1},\ldots,\yy_u), ~~~~for ~ i \in N_1,\\
  &\xx_i=l_{u+v+i}(\xx_{i+1},\ldots,\zz_d), ~for ~ i \in N_3.
\end{align*}
Also, $h_2=0 $ implies that
\begin{align*}
  &\zz_i=l_{u+i}(\xx,\zz_{i+1},\ldots,\zz_v), ~for ~ i \in N_2,\\
  &\xx_i=l_{u+v+i}(\xx_{i+1},\ldots,\zz_d), ~for ~ i \in N_3.
\end{align*}
Now, let \oomit{Let divide each of $\yy,\zz,\xx$ into two parts by $N_1,N_2,N_3$}
\begin{align*}
 & \yy^1 = (y_{i_1},\ldots, y_{i_{|N_1|}}),
         \yy^2= (y_{j_1}, \ldots, y_{j_{u-|N_1|}}),  \\
          &   \quad \quad \mbox{ where }
           \{i_1,\ldots, i_{|N_1|}\} =N_1, \{j_1,\ldots,j_{u-|N_1|}\} = \{1,\ldots, u\} -N_1,\\
 & \zz^1 = (z_{i_1},\ldots, z_{i_{|N_2|}}),
         \zz^2= (z_{j_1}, \ldots, z_{j_{u-|N_2|}}),  \\
          &   \quad \quad \mbox{ where }
           \{i_1,\ldots, i_{|N_2|}\} =N_2, \{j_1,\ldots,j_{v-|N_2|}\} = \{1,\ldots, v\} -N_2,\\
 & \xx^1 = (x_{i_1},\ldots, x_{i_{|N_3|}}),
         \xx^2= (x_{j_1}, \ldots, x_{j_{u-|N_3|}}),  \\
          &   \quad \quad \mbox{ where }
           \{i_1,\ldots, i_{|N_3|}\} =N_3, \{j_1,\ldots,j_{d-|N_3|}\} = \{1,\ldots, d\} -N_3.
\end{align*}
Clearly, $\#(\textit{Var}(\xx^1)+\textit{Var}(\yy^1)+\textit{Var}(\zz^1)) > 0$.
By  linear algebra, there exist three matrices $\Lambda_1,\Lambda_2,\Lambda_3$ and three vectors $\gamma_1,\gamma_2,\gamma_3$ s.t.
\begin{align*}
&\yy^1 = \Lambda_1 \left( \begin{matrix} \xx^2 \\ \yy^2 \end{matrix} \right) + \gamma_1,\\
&\zz^1 = \Lambda_2 \left( \begin{matrix} \xx^2 \\ \zz^2 \end{matrix} \right) + \gamma_2,\\
&\xx^1 = \Lambda_3 \xx^2 + \gamma_3.
\end{align*}
Since
  $\phi_2 \models h_1 = 0, ~~~~ \psi_2 \models h_2=0$,
then,
\begin{align*}
  &\phi_2 \models ( (\yy^1 = \Lambda_1 \left( \begin{matrix} \xx^2 \\ \yy^2 \end{matrix} \right) + \gamma_1)\wedge (\xx^1 = \Lambda_3 \xx^2 + \gamma_3) ), \\
  &\psi_2 \models  ((\zz^1 = \Lambda_2 \left( \begin{matrix} \xx^2 \\ \zz^2 \end{matrix} \right) + \gamma_2)\wedge (\xx^1 = \Lambda_3 \xx^2 + \gamma_3) ).
\end{align*}
\qed
\end{proof}

So, replacing $(\xx^1,\yy^1)$ in $f_i(\xx,\yy)$ and $g_j(\xx,\yy)$ by $\Lambda_3 \xx^2 + \gamma_3$
 $\Lambda_1 \left( \begin{matrix} \xx^2 \\ \yy^2 \end{matrix} \right) + \gamma_1$ respectively,
  results in new polynomials $\hat{f_i}(\xx^2,\yy^2)$ and $\hat{g_j}(\xx^2,\yy^2)$, for $i=1,\ldots,r_1$, $j=1,\ldots,s_1$.
Similarly, replacing $(\xx^1,\zz^1)$ in $f_i(\xx,\zz)$ and $g_j(\xx,\zz)$ by $ \Lambda_3 \xx^2 + \gamma_3$ and
 $\Lambda_2 \left( \begin{matrix} \xx^2 \\ \zz^2 \end{matrix} \right) + \gamma_2$ respectively, derives new polynomials $\hat{f_i}(\xx^2,\zz^2)$ and $\hat{g_j}(\xx^2,\zz^2)$, for $i=r_1+1,\ldots,r$, $j=s_1+1,\ldots,s$.  Regarding the resulted polynomials above, we have the following property.
\begin{lemma} \label{concave-hold}
Let $\xi \in \RR^m$ and $\zeta \in \RR^n$ be two vector variables,
$g(\xi,\zeta)= \left( \begin{matrix} &\xi \\
&\zeta \end{matrix} \right)^T G \left( \begin{matrix} &\xi \\
&\zeta \end{matrix} \right) +  a^T \left( \begin{matrix} &\xi \\
&\zeta \end{matrix} \right) + \alpha$ be a CQ polynomial on $(\xi,\zeta)$,
i.e. $G \preceq 0$.  Replacing $\zeta$ in $g$ by $\Lambda \xi + \gamma$ derives
$\hat{g}(\xi) = g(\xi, \Lambda \xi + \gamma)$, then $\hat{g}(\xi)$ is a CQ
polynomial in $\xi$.
\end{lemma}
\begin{proof}
  $G \preceq 0$ iff $- \left( \begin{matrix} &\xi \\
&\zeta \end{matrix} \right)^T G \left( \begin{matrix} &\xi \\
&\zeta \end{matrix} \right)$ is an SOS. Thus, there exist $l_{i,1} \in \RR^m$,
$l_{i,2} \in \RR^n$, for
$i=1,\ldots, s$, $s \in \NN^{+}$ s.t.
 $\left( \begin{matrix} &\xi \\
&\zeta \end{matrix} \right)^T G \left( \begin{matrix} &\xi \\
&\zeta \end{matrix} \right) = - \sum_{i=1}^s (l_{i,1}^T \xi + l_{i,2}^T \zeta)^2$.
Hence,
\begin{align*}
 \left( \begin{matrix} &\xi \\
&\Lambda \xi + \gamma
\end{matrix} \right)^T G \left( \begin{matrix} &\xi \\
&\Lambda \xi + \gamma
\end{matrix} \right) = - \sum_{i=1}^s (l_{i,1}^T \xi + l_{i,2}^T (\Lambda \xi + \gamma))^2\\
=- \sum_{i=1}^s ((l_{i,1}^T + l_{i,2}^T \Lambda) \xi + l_{i,2}^T \gamma)^2 \\
=- \sum_{i=1}^s ((l_{i,1}^T + l_{i,2}^T \Lambda) \xi)^2 + l(\xi),
\end{align*}
where $l(\xi)$ is a linear function in $\xi$.
Then we have
\begin{align*}
  \hat{g}(\xi) = - \sum_{i=1}^s ((l_{i,1}^T + l_{i,2}^T \Lambda) \xi)^2 + l(\xi)+
  \va^T \left( \begin{matrix} &\xi \\
&\Lambda \xi + \gamma
\end{matrix} \right) + \alpha.
\end{align*}
Obviously, there exist $\hat{G} \preceq 0$, $\hat{\va}$ and $\hat{\alpha}$ such that
\begin{align*}
  \hat{g} = \xi \hat{G} \xi^T + \hat{\va}^T \xi + \hat{\alpha}.
\end{align*}
Therefore, $\hat{g}$ is concave quadratic polynomial in $\xi$. \qed
\end{proof}

\begin{theorem} \label{the:gcase:2}
  In the proof of Lemma \ref{lemma:dec}, if $a_{u+v+d+1} = 0$, then
  Lemma \ref{lemma:elim} holds. So, let $\hat{f_i}$ and $\hat{g_j}$ as above,
  and
\begin{align*}
  &\phi' = \bigwedge_{i=1}^{r_1} \hat{f_i} \ge 0 \wedge
  \bigwedge_{j=1}^{s_1} \hat{g_j} >0, \\
  &\psi' = \bigwedge_{i=r_1+1}^{r} \hat{f_i} \ge 0 \wedge
  \bigwedge_{j=s_1+1}^{s} \hat{g_j} >0.
\end{align*}
Then $\phi'$ and $\psi'$ satisfy $(i)-(iv)$.
\end{theorem}
\begin{proof}
From Lemma \ref{lemma:elim}, we have
\begin{align*}
  &\phi_2 \models ( (\yy^1 = \Lambda_1 \left( \begin{matrix} \xx^2 \\ \yy^2 \end{matrix} \right) + \gamma_1)\wedge (\xx^1 = \Lambda_3 \xx^2 + \gamma_3) ), \\
  &\psi_2 \models  ((\zz^1 = \Lambda_2 \left( \begin{matrix} \xx^2 \\ \zz^2 \end{matrix} \right) + \gamma_2)\wedge (\xx^1 = \Lambda_3 \xx^2 + \gamma_3) ).\\
\end{align*}
Let
\begin{align*}
  &\phi_2' := ( (\yy^1 = \Lambda_1 \left( \begin{matrix} \xx^2 \\ \yy^2 \end{matrix} \right) + \gamma_1)\wedge (\xx^1 = \Lambda_3 \xx^2 + \gamma_3) \wedge \phi ), \\
  &\psi_2' :=  ((\zz^1 = \Lambda_2 \left( \begin{matrix} \xx^2 \\ \zz^2 \end{matrix} \right) + \gamma_2)\wedge (\xx^1 = \Lambda_3 \xx^2 + \gamma_3) \wedge \psi ).\\
\end{align*}
Then $\phi_2 \models \phi_2'$, $\phi_2 \models \phi_2'$ and $\phi_2'\wedge\psi_2'\models\bot$. Thus any interpolant for $\phi_2'$ and $\psi_2'$ is also an interpolant of $\phi_2$ and $\psi_2$.

By the definition of $\phi'$ and $\psi'$, it follows  $\phi' \wedge \psi' \models \bot$ iff  $\phi_2^{'}\wedge\psi_2^{'}\models\bot$,
so $\phi' \wedge \psi' \models \bot$, $(i)$ holds.

Moreover, $\phi_2{'}\models \phi'$, $\psi_2{'}\models \psi'$, $\textit{Var}(\phi') \subseteq \textit{Var}(\phi_2{'})$ and $\textit{Var}(\psi') \subseteq \textit{Var}(\psi_2{'})$, then any interpolant for
$\phi'$ and $\psi'$ is also an interpolant for $\phi_2{'}$ and $\psi_2{'}$, then also an
interpolant for $\phi_2$ and $\psi_2$. By Theorem \ref{lemma:p22}, $(iii)$ holds.

Since $\#(\textit{Var}(\phi)+\textit{Var}(\psi)) - \#(\textit{Var}(\phi')+\textit{Var}(\psi'))=\#(\xx^1,\yy^1,\zz^1) >0$,
then $(vi)$ holds.

For $(ii)$, $\phi',\psi'$ have the same form with $\phi,\psi$, means that
$\hat{f_i}, i=1,\ldots,r$ are CQ and $\hat{g_j}, j=1,\ldots,s$
are CQ. This is satisfied directly by
Lemma \ref{concave-hold}.
\qed
\end{proof}

The following simple example illustrates how the above
construction works.

\begin{example} \label{exam1}
  Let $f_1 = x_1, f_2 = x_2,f_3= -x_1^2-x_2^2 -2x_2-z^2, g_1= -x_1^2+2 x_1 - x_2^2 + 2 x_2 - y^2$. Two formulas $\phi:=(f_1 \ge 0) \wedge (f_2 \ge0) \wedge (g_1 >0)$,
  $\psi := (f_3 \ge 0)$. $\phi \wedge \psi \models \bot$.

  The condition $\sosc$ does not hold, since
  \begin{align*}
    -(0 f_1 + 2 f_2 + f_3) = x_1^2 +x_2^2 + z^2 {\rm ~ is ~ a ~ sum ~ of ~ square}.
  \end{align*}
  Then we have $h=x_1^2 +x_2^2 + z^2$, and
  \begin{align}
    h_1 = \frac{1}{2}x_1^2+\frac{1}{2}x_2^2,~~ h_2 =\frac{1}{2}x_1^2+ \frac{1}{2}x_2^2 + z^2.  \label{h:choose}
  \end{align}
  Let $f = 0 f_1 + 2 f_2 + h_1 =
  \frac{1}{2}x_1^2+\frac{1}{2}x_2^2+2x_2$.

For the recursive call,  we have $f = 0$ as well as $x_1 = 0, x_2
= 0$ from $h_1=0$ to construct
  $\phi'$ from $\phi$; similarly $\psi'$ is constructing
  by setting $x_1=x_2=0,z=0$ in $\psi$ as derived from $h_2 = 0$.
  \begin{align*}
    \phi' =0 \ge 0 \wedge 0 \ge 0 \wedge -y^2 > 0 = \bot, ~~\psi^{'} = 0 \ge 0 = \top.
  \end{align*}
  Thus, $I(\phi',\psi'):=(0 > 0)$ is an interpolant for $(\phi',\psi')$.

  An interpolant for $\phi$ and $\psi$ is thus
  $(f(x) >0 ) \vee (f(x)=0 \wedge I(\phi',\psi'))$, which is
  $\frac{1}{2}x_1^2+\frac{1}{2}x_2^2+2x_2 > 0$.
\end{example}

\oomit{ By Theorem \ref{the:gcase:1} and Theorem \ref{the:gcase:2}, when
 $\textit{Var}(\phi) \nsubseteq \textit{Var}(\psi)$ and the
 condition $\sosc$ does not hold, we can solve Problem 1 in a
 recursive way. From $(vi)$ we know that this recursion must
 terminate at most $d+u+v$ times. If it terminates at
 $\phi',\psi'$ with $\sosc$, then Problem 1 is solved by Theorem
 \ref{the:int}; otherwise, it terminates at $\phi',\psi'$ with
 $\textit{Var}(\phi') \subseteq \textit{Var}(\psi')$, then
 $\phi'$ itself is an interpolant for $\phi'$ and $\psi'$.
}

\subsection{Algorithms} \label{sec:alg}

Algorithm $\igfch$ deals with the case when $\phi$ and $\psi$
satisfy the $\sosc$ condition.
\begin{algorithm}[!htb]
	\label{alg:int}
	\SetKwData{Left}{left}\SetKwData{This}{this}\SetKwData{Up}{up}
	\SetKwFunction{Union}{Union}\SetKwFunction{FindCompress}{FindCompress}
	\SetKwInOut{Input}{input}\SetKwInOut{Output}{output}
    \caption{ {\tt $\igfch$ }}
	\Input{Two formulas $\phi$, $\psi$ with $\sosc$ and $\phi \wedge \psi \models \bot$, where
$\phi= f_1 \ge 0 \wedge \ldots \wedge f_{r_1} \ge 0 \wedge g_1 >0 \wedge \ldots \wedge g_{s_1} > 0 $,
$\psi= f_{r_1+1} \ge 0 \wedge \ldots \wedge f_{r} \ge 0 \wedge g_{s_1+1} >0 \wedge \ldots \wedge g_{s} > 0 $,
$f_1, \ldots, f_{r}, g_1, \ldots, g_s$ are all concave quadratic polynomials,
$f_1, \ldots, f_{r_1}, g_1, \ldots, g_{s_1} \in \RR[\xx,\yy]$,
$f_{r_1+1}, \ldots, f_{r}, g_{s_1+1}, \ldots, g_{s} \in \RR[\xx,\zz]$
}
	\Output{A formula $I$ to be a Craig interpolant for $\phi$  and $\psi$}
	\SetAlgoLined
	\BlankLine
    \textbf{Find} $\lambda_1,\ldots,\lambda_r \ge 0,\eta_0,\eta_1,\ldots,\eta_s \ge 0, h_1 \in \RR[\xx,\yy], h_2 \in \RR[\xx,\zz]$ by SDP s.t.
    \begin{align*}
    & \sum_{i=1}^{r} \lambda_i g_j+\sum_{j=1}^{s} \eta_j g_j + \eta_0 + h_1+h_2 \equiv 0,\\
    & \eta_0+\eta_1 + \ldots + \eta_s = 1,\\
    & h_1, h_2 {\rm ~ are ~ SOS~polynomial};
    \end{align*}\\
    \tcc{This is essentially a $\sdp$ problem, see Section \ref{sec:hold}}
    $f:=\sum_{i=1}^{r_1} \lambda_i g_j+\sum_{j=1}^{s_1} \eta_j g_j + \eta_0 + h_1$\;
    \textbf{if } $\sum_{j=0}^{s_1} \eta_j > 0$ \textbf{ then } $I:=(f>0)$;
     \textbf{else} $I:=(f\ge 0)$\;
    \KwRet  $I$ \label{subalg:return}
\end{algorithm}

\begin{theorem}[Soundness and Completeness of $\igfch$] \label{thm:correctness-1}
  $\igfch$ computes an interpolant $I$ of mutually contradictory $\phi, \psi$ with CQ
  polynomial inequalities satisfying the
  $\sosc$ condition .
\end{theorem}
\begin{proof}
  It is guaranteed by Theorem \ref{the:int}. \qed
\end{proof}

The recursive algorithm  $\igfch$ is given below. For the base
case when $\phi, \psi$ satisfy the $\sosc$ condition, it invokes $\igfch$.

\begin{algorithm}[!htb]
	\label{alg:int}
	\SetKwData{Left}{left}\SetKwData{This}{this}\SetKwData{Up}{up}
	\SetKwFunction{Union}{Union}\SetKwFunction{FindCompress}{FindCompress}
	\SetKwInOut{Input}{input}\SetKwInOut{Output}{output}
    \caption{ {\tt $\igfqc$ }\label{prob:in-out}}
	\Input{Two formulas $\phi$, $\psi$ with $\phi \wedge \psi \models \bot$, where
$\phi= f_1 \ge 0 \wedge \ldots \wedge f_{r_1} \ge 0 \wedge g_1 >0 \wedge \ldots \wedge g_{s_1} > 0 $,
$\psi= f_{r_1+1} \ge 0 \wedge \ldots \wedge f_{r} \ge 0 \wedge g_{s_1+1} >0 \wedge \ldots \wedge g_{s} > 0 $,
$f_1, \ldots, f_{r}, g_1, \ldots, g_s$ are all CQ polynomials,
$f_1, \ldots, f_{r_1}, g_1, \ldots, g_{s_1} \in \RR[\xx,\yy]$, and
$f_{r_1+1}, \ldots, f_{r}, g_{s_1+1}, \ldots, g_{s} \in \RR[\xx,\zz]$
}
	\Output{A formula $I$ to be a Craig interpolant for $\phi$ and $\psi$}
	\SetAlgoLined
	\BlankLine
    \textbf{if} $\textit{Var}(\phi)\subseteq \textit{Var}(\psi)$ \textbf{then} $I:=\phi$; \KwRet $I$\; \label{alg2:1}
     \textbf{Find} $\delta_1,\ldots,\delta_r \ge 0, h \in \RR[\xx,\yy,\zz]$ by SDP s.t.  $\sum_{i=1}^r \delta_i f_i +h \equiv 0$ and $h$ is
    SOS; \label{alg2:2}\\
	\tcc{Check the condition $\sosc$}
	\textbf{if} \emph{no solution} \textbf{then} $I := \igfch(\phi, \psi)$; \label{cond:hold}
     \KwRet $I$\; \label{alg2:3}
    \tcc{ $\sosc$ holds}
    Construct $h_1 \in \RR[\xx,\yy]$ and $h_2 \in \RR[\xx,\zz]$ with the forms $\mathrm{(H1)}$ and $\mathrm{(H2)}$\;
    \label{alg2:4}
    $f:=\sum_{i=1}^{r_1} \delta_i f_i +h_1 =-\sum_{i=r_1}^{r} \delta_i f_i -h_2  $\; \label{alg2:5}
    Construct
    $\phi'$ and $\psi'$ using Theorem \ref{the:gcase:1} and Theorem
    \ref{the:gcase:2} by eliminating variables due to
    $h_1 = h_2 = 0$\; \label{alg2:6}
     $I' = \igfqc(\phi', \psi')$\;     \label{alg2:7}
    $I:=(f>0) \vee (f \ge 0 \wedge I')$\; \label{alg2:8}
    \KwRet $I$ \label{alg2:9}
\end{algorithm}

\begin{theorem}[Soundness and Completeness of $\igfqc$] \label{thm:correctness-2}
 $\igfqc$  computes an interpolant $I$ of mutually contradictory $\phi, \psi$ with CQ
 polynomial inequalities.
\end{theorem}
\begin{proof}
  If $\textit{Var}(\phi) \subseteq \textit{Var}(\psi)$, $\igfqc$ terminates  at step \ref{alg2:1}, and
  returns $\phi$ as an interpolant. Otherwise, there are two cases:

  (i) If $\sosc$ holds, then $\igfqc$ terminates at step \ref{alg2:3} and
  returns an interpolant for $\phi$ and $\psi$ by calling
  $\igfch$. Its soundness and completeness follows from the
  previous theorem.

  (ii)  $\textit{Var}(\phi) \nsubseteq \textit{Var}(\psi)$ and
  $\sosc$ does not hold: The proof is by induction on the number
  of recursive calls to $\igfqc$, with the case of 0 recursive
  calls being (i) above.

 In the induction step, assume that for a $k^{th}$-recursive call to
 $\igfqc$ gives a correct interpolant $I'$ for $\phi'$ and
$\psi'$, where $\phi'$ and $\psi'$ are constructed by Theorem \ref{the:gcase:1} or
  Theorem \ref{the:gcase:2}.

  By Theorem \ref{the:gcase:2},
  the interpolant $I$ constructed from $I'$ is the correct answer
 for $\phi$ and $\psi$.

  The recursive algorithm terminates in all three cases: (i)
  $\textit{Var}(\phi) \subseteq \textit{Var}(\psi)$, (ii) $\sosc$
  holds, which is achieved at most $u+v+d$ times by Theorem
  \ref{the:gcase:2}, and (iii) the number of variables in $\phi',
  \psi'$ in the recursive call is smaller than the number of
  variables in $\phi, \psi$.

\oomit{Meanwhile, for these two basic cases,
    we have already know that this algorithm return the right
    answer, by inductive method, the algorithm return the right
    answer with input $\phi$ and $\psi$.}  \qed
\end{proof}

\subsection{Complexity analysis of $\igfch$ and $\igfqc$}
It is well known that an $\sdp$ problem
can be solved in polynomial time complexity. We analyze the
complexity of the above algorithms assuming that
the complexity of an $\sdp$ problem is of time complexity $g(k)$,
where $k$ is the input size.

\begin{theorem} \label{thm:complexity-1}
  The complexity of $\igfch$ is $\mathcal{O}(g(r+s+n^2))$, where $r$
  is the number of nonstrict inequalities $f_i$s and $s$ is the
  number of strict inequalities $g_j$s, and $n$
  is the number of variables in $f_i$s and $g_j$s.
\end{theorem}
\begin{proof}
  In this algorithm we first need to solve a constraint solving problem in step $1$, see Section
  \ref{sec:hold}, it is an $\sdp$ problem with size $\mathcal{O}(r+s+n^2)$, so the complexity
  of step $1$ is $\mathcal{O}(g(r+s+n^2))$. Obviously,  the complexity of steps
  $2-4$ is linear in $(r+s+n^2)$, so the complexity of $\igfch$ is $\mathcal{O}(g(r+s+n^2))$. \qed
\end{proof}

\begin{theorem} \label{thm:complexity-2}
  The complexity of $\igfqc$ is $\mathcal{O}(n* g(r+s+n^2) )$,
  where $r, s, n$ are as defined in the previous theorem.
\end{theorem}
\begin{proof}
  The algorithm $\igfqc$ is a recursive algorithm, which is called
   at most $n$ times, since in every recursive call, at least one
   variable gets eliminated. Finally, it terminates at step $1$ or step
  $3$ with complexity $\mathcal{O}(g(r+s+n^2))$.

 The complexity of each recursive call, i.e., the complexity
  for step $2$ and steps $4-9$, can be analyzed as follows:

  For step $2$, checking if  $\sosc$ holds is done by solving the following problem: \\
\textbf{ find: }
  $\delta_1,\ldots,\delta_r \ge 0$, and an
  SOS polynomial $ h \in \RR[\xx,\yy,\zz]$
s.t.
 $\sum_{i=1}^r \delta_i f_i +h \equiv 0$,

\noindent which  is equivalent to the following linear matrix inequality ($\lmi$), \\
{\bf find: }
  $\delta_1,\ldots,\delta_r \ge 0$, $M \in R^{(n+1 \times (n+1)}$,
s.t.
 $M=-\sum_{i=1}^r \delta_i P_i$, $M\succeq  0$,
where $P_i \in R^{(n+1) \times (n+1)}$ is defined as (\ref{pq:def}).
Clearly,  this is an $\sdp$ problem with size $\mathcal{O}(r+n^2)$, so the complexity of this step  is $\mathcal{O}( g(r+n^2) )$.

For steps $4-9$, by the proof of Lemma \ref{h:sep}, it is easy to see that to represent $h$ in the
form $\mathrm{(H)}$ in Lemma \ref{lem:split} 
can be done with complexity
$\mathcal{O}(n^2)$,
$h_1$ and $h_2$ can be computed with complexity $\mathcal{O}(n^2)$.  Thus,
the complexity of step $4$ is $\mathcal{O}(n^2)$. Step $5$ is much easy. For step $6$,
  using linear algebra operations, it is
easy to see that the complexity is $\mathcal{O}(n^2+r+s)$.
So,  the complexity is $\mathcal{O}(n^2+r+s)$ for step $4-9$.

In a word, the overall complexity of $\igfqc$ is
\begin{eqnarray*}
 \mathcal{O}(g(r+s+n^2))+n \mathcal{O}(n^2+r+s)
  & = & \mathcal{O}(n * g(r+s+n^2) ).
\end{eqnarray*}
\qed
\end{proof}

\section{Combination:  quadratic concave polynomial inequalities
  with uninterpreted function symbols (\textit{EUF})}

This section combines the quantifier-free theory of quadratic
concave polynomial inequalities with the theory of equality over
uninterpreted function symbols (\textit{EUF}).
\oomit{Using hierarchical
reasoning framework proposed in \cite{SSLMCS2008} which was
applied in \cite{RS10} to generate interpolants for mutually
contradictory formulas in the combined quantfier-free theory of linear
inequalities over the reals and equality over uninterpreted
symbols, we show below how the algorithm $\igfqc$ for quadratic concave
polynomial inequalities over the reals can be extended to
generate interpolants for mutually contradictory formulas
consisting of quadratic concave polynomials expressed using terms
built from unintepreted symbols.}
The proposed algorithm for generating interpolants for the combined
theories is presented in Algorithm~\ref{alg:euf}. As the reader would observe,
it is patterned after the algorithm $\text{INTER}_{LI(Q)^\Sigma}$ in Figure 4 in
\cite{RS10} following the hierarchical reasoning and
interpolation generation framework in \cite{SSLMCS2008}  with the following key differences\footnote{The
  proposed algorithm andd its way of handling of combined theories
  do not crucially depend upon using algorithms in \cite{RS10};
  however, adopting their approach makes proofs and presentation
  easier by focusing totally on the quantifier-free theory of CQ  polynomial
  inequalities.}:
\begin{enumerate}
\item To generate interpolants for mutually contradictory
  conjunctions of CQ polynomial
  inequalities, we call $\igfqc$.
\item We prove below that (i) a nonlinear equality over
  polynomials cannnot be generated from CQ
  polynomials, and furthermore (ii) in the base case when the $\sosc$
  condition is satisfied by CQ polynomial
  inequalities, linear equalities are deduced only from the linear
  inequalities in a problem (i.e., nonlinear inequalities do not play any
  role); separating terms for mixed equalities are computed the
  same way as in the algorithm SEP in \cite{RS10},  and (iii) as shown in Lemmas \ref{h:sep},
  \ref{lem:split} and Theorem \ref{the:gcase:2}, during recursive calls to $\igfqc$, additional
  linear unmixed equalities are deduced which are local to either $\phi$
  or $\psi$, we can use these equalities as well as those in
  (ii) for the base case to reduce the number of variables
  appearing in $\phi$ and $\psi$ thus reducing the complexity of
  the algorithm; 
  { equalities relating variables of $\phi$ are also
  included in the interpolant}.
\end{enumerate}
Other than that, the proposed algorithm reduces to
$\text{INTER}_{LI(Q)^\Sigma}$ if $\phi, \psi$ are purely from $LI(Q)$ and/or
$\textit{EUF}$.

In order to get directly to the key concepts used, we assume the reader's
familiarity with the basic construction of flattening and
purification by introducing fresh variables for the arguments
containing uninterpreted functions.

\oomit{\begin{definition}
  Given two formulas $\phi$ and $\psi$ with $\phi \wedge \psi \models \bot$. A formula $I$ is said to be
  an interpolant for $\phi$ and $\psi$, if the following three conditions hold, $(i)$ $\phi \models I$, $(ii)$
  $I \wedge \psi \models \bot$, and $(iii)$ the variables and function symbols in $I$ is in both $\phi$ and $\psi$,
  i.e., $\textit{Var}(I) \subset \textit{Var}(\phi) \cap \textit{Var}(\psi) \wedge FS(I) \subset FS(\phi) \cap FS(\psi)$, where $FS(w)$ means that
  the function symbols in $w$.
\end{definition} }

\subsection{Problem Formulation}

Let $\Omega = \Omega_1 \cup \Omega_2 \cup \Omega_3$ be a finite
set of uninterpreted function symbols in $\textit{EUF};$ further, denote
$\Omega_1 \cup \Omega_2$ by $\Omega_{12}$
and $\Omega_1 \cup \Omega_3$ by $\Omega_{13}$.
Let $\RR[\xx,\yy,\zz]^{\Omega}$ be the extension of $\RR[\xx,\yy,\zz]$
in which polynomials can have terms built using function symbols
in $\Omega$ and variables in $\xx, \yy, \zz$.

\begin{problem} \label{EUF-problem}
Suppose two formulas $\phi$ and $\psi$ with
$\phi \wedge \psi \models \bot$, where
$\phi= f_1 \ge 0 \wedge \ldots \wedge f_{r_1} \ge 0 \wedge g_1 >0
\wedge \ldots \wedge g_{s_1} > 0 $,
$\psi= f_{r_1+1} \ge 0 \wedge \ldots \wedge f_{r} \ge 0 \wedge
g_{s_1+1} >0 \wedge \ldots \wedge g_{s} > 0 $,
where $f_1, \ldots, f_{r}, g_1, \ldots, g_s$ are all CQ polynomial,
$f_1, \ldots, f_{r_1}, g_1, \ldots, g_{s_1} \in
\RR[\xx,\yy]^{\Omega_{12}}$,
$f_{r_1+1}, \ldots, f_{r}, g_{s_1+1}, \ldots, g_{s} \in
\RR[\xx,\zz]^{\Omega_{13}}$,
the goal is to generate an
interpolant $I$ for $\phi$ and $\psi$,  expressed using the common
symbols $\xx, \Omega_1$, i.e., $I$ includes only polynomials in $\RR[\xx]^{\Omega_1}$.
\end{problem}

{\bf Flatten and Purify:} Purify and flatten the formulas $\phi$
and $\psi$ by introducing fresh variables for each term with
uninterpreted symbols as well as for the terms with uninterpreted
symbols. Keep track of new variables introduced exclusively for
$\phi$ and $\psi$ as well as new common variables.

 Let $\overline{\phi} \wedge \overline{\psi} \wedge \bigwedge D$
 be obtained from $\phi \wedge \psi$ by flattening and purification
 where $D$ consists of unit clauses of the form
 $\omega(c_1,\ldots,c_n)=c$, where $c_1,\ldots,c_n$ are variables
 and $\omega \in \Omega$.
Following \cite{SSLMCS2008,RS10}, using the axiom of
an uninterpreted function symbol, a set $N$ of Horn clauses are generated as follows,
$$
N=\{ \bigwedge_{k=1}^n c_k=b_k \rightarrow c=b \mid \omega(c_1,\ldots,c_n)=c \in D, \omega(b_1,\ldots,b_n)=b \in D \}.
$$
The set $N$ is partitioned into $N_{\phi}, N_{\psi}, N_{\text{mix}}$
with all symbols in $N_{\phi}, N_{\psi}$ appearing in $\overline{\phi}$, $\overline{\psi}$,
respectively, and $N_{\text{mix}}$ consisting of symbols from both $\overline{\phi}, \overline{\psi}$.

It is easy to see that for every Horn clause in $N_{\text{mix}}$, each of
equalities in the hypothesis as well as conclusion is mixed.

\begin{eqnarray} \label{eq:reducedP}
  \phi \wedge \psi \models \bot \mbox{ iff } \overline{\phi} \wedge \overline{\psi} \wedge D \models \bot
  \mbox{ iff } (\overline{\phi}\wedge N_{\phi}) \wedge (\overline{\psi}  \wedge N_{\psi}) \wedge N_{\text{mix}} \models \bot.
\end{eqnarray}

Notice that $ \overline{\phi} \wedge \overline{\psi}  \wedge N
\models \bot$ has no uninterpreted function symbols. An interpolant generated
for this problem\footnote{after properly handling
  $N_{\text{mix}}$ since Horn clauses have symbols both from
$\overline{\phi}$ and $\overline{\psi}$.} can be used to
generate an interpolant for $\phi, \psi$ after uniformly
replacing all new symbols by their corresponding expressions from $D$.

\subsection{Combination algorithm}
If $N_{\text{mix}}$ is empty, implying there are
no mixed Horn clauses, then the algorithm invokes $\igfqc$ on a
finite set of subproblems generated from a disjunction of
conjunction of polynomial inequalities obtained after expanding Horn
clauses in $N_{\phi}$ and $N_\psi$ and applying De Morgan's
rules. The resulting interpolant is a disjunction of the
interpolants generated for each subproblem.

The case when $N_{\text{mix}}$ is nonempty is more interesting, but it has the same structure as
the algorithm $\text{INTER}_{LI(Q)^\Sigma}$ in \cite{RS10} except that
instead of $\text{INTER}_{LI(Q)}$, it calls $\igfqc$.

The following lemma proves that if a conjunction of polynomial
inequalities satisfies the $\sosc$ condition and an
equality on variables can be deduced from it, then it suffices to
consider only linear inequalities in the conjunction. This
property enables us to use algorithms used in \cite{RS10} to
generate such equalities as well as separating terms for the
constants appearing in mixed equalities (algorithm SEP in
\cite{RS10}).

\begin{lemma} \label{lemma:qc2lin}
Let $f_i$, $i=1,\ldots,r$ be CQ polynomials, and $\lambda_i \ge 0$, if
     $\sum_{i=1}^{r} \lambda_i f_i   \equiv 0$,
  then for any $1\le i \le r$,
    $\lambda_i=0$  or $f_i$  is linear.
\end{lemma}

\begin{proof}
  Let $f_i = \xx^T A_i \xx + l_i^T \xx + \gamma_i$, then $A_i \preceq 0$, for $i=1, \ldots, r$.
  Since $\sum_{i=1}^{r} \lambda_i f_i = 0$, we have $\sum_{i=1}^{r} \lambda_i A_i   = 0$.
  Thus for any $1\le i \le r$, $\lambda_i=0$ or $A_i=0$. \qed
\end{proof}

\begin{lemma} \label{lem:linear-part}
  Let $\overline{\phi}$ and $\overline{\psi}$ be obtained as above with  $\sosc$. If $\overline{\phi} \wedge \overline{\psi}$ is satisfiable, $\overline{\phi} \wedge \overline{\psi} \models c_k=b_k$, then
    $LP(\overline{\phi}) \wedge LP(\overline{\psi}) \models c_k=b_k$,
  where $LP(\overline{\phi})$  ($LP(\overline{\psi})$) is a formula defined by all the linear constraints in
  $\overline{\phi}$ ($\overline{\psi}$).
\end{lemma}

\begin{proof}
  Since $\overline{\phi} \wedge \overline{\psi} \models c_k=b_k$, then $\overline{\phi} \wedge \overline{\psi} \wedge c_k>b_k \models \bot$. By Theorem \ref{the:int},
  there exist $\lambda_i\ge 0$ ($i=1,\cdots,r$), $\eta_j \ge 0$ ($j=0,1,\cdots,s$), $\eta \ge 0$ and two quadratic SOS polynomials $\overline{h}_1 $ and $\overline{h}_2 $ such that
  \begin{align}
    & \sum_{i=1}^{r} \lambda_i \overline{f}_i +\sum_{j=1}^{s} \eta_j \overline{g}_j+\eta (c_k-b_k) + \eta_0 + \overline{h}_1+\overline{h}_2 \equiv 0, \label{cond:sep:1}\\
    & \eta_0+\eta_1 + \ldots + \eta_s +\eta= 1.\label{cond:sep:2}
  \end{align}
   As $\overline{\phi} \wedge \overline{\psi}$ is satisfiable and $\overline{\phi} \wedge \overline{\psi} \models c_k=b_k$, there exist $\xx_0,\yy_0,\zz_0,\aa_0,\bb_0,\cc_0$ s.t.
     $\overline{\phi}[\xx/\xx_0, \yy/\yy_0,\aa/\aa_0,\cc/\cc_0]$, $\overline{\psi}[\xx/\xx_0, \zz/\zz_0,\bb/\bb_0,\cc/\cc_0]$,
  and $c_k=b_k[\aa/\aa_0,\bb/\bb_0,\cc/\cc_0]$. Thus, it follows that
  $\eta_0=\eta_1=\ldots=\eta_s=0$ from (\ref{cond:sep:1}) and $\eta=1$  from (\ref{cond:sep:2}).
  Hence, (\ref{cond:sep:1}) is equivalent to
  \begin{align} \label{amb}
     \sum_{i=1}^{r} \lambda_i \overline{f}_i + (c_k-b_k) + \overline{h}_1+\overline{h}_2 \equiv 0.
  \end{align}
  Similarly, we can prove that there exist $\lambda_i'\ge 0$ ($i=1,\cdots,r$) and two quadratic SOS polynomials $h_1'$ and $h_2'$ such that
  \begin{align} \label{bma}
     \sum_{i=1}^{r} \lambda_i' \overline{f}_i + (b_k-c_k) + \overline{h}_1'+\overline{h}_2' \equiv 0.
  \end{align}
  From (\ref{amb}) and (\ref{bma}), it follows
  \begin{align} \label{bmap}
     \sum_{i=1}^{r} (\lambda+\lambda_i') \overline{f}_i  + \overline{h}_1+\overline{h}_1'+\overline{h}_2+\overline{h}_2' \equiv 0.
  \end{align}
  In addition, $\sosc$ implies  $\overline{h}_1\equiv \overline{h}_1' \equiv \overline{h}_2 \equiv \overline{h}_2' \equiv0$. So
  \begin{align} \label{amb1}
     \sum_{i=1}^{r} \lambda_i \overline{f}_i + (c_k-b_k)  \equiv 0,
  \end{align}
  and
  \begin{align} \label{bma1}
     \sum_{i=1}^{r} \lambda_i' \overline{f}_i + (b_k-c_k)  \equiv 0.
  \end{align}
  Applying  Lemma \ref{lemma:qc2lin} to (\ref{amb1}), we have that
   $\lambda_i=0$ or $f_i$ is linear.
  So
  \begin{align*}
    LP(\overline{\phi}) \wedge LP(\overline{\psi}) \models c_k \le b_k.
  \end{align*}
  Likewise, by applying Lemma \ref{lemma:qc2lin} to (\ref{bma1}), we have
  \begin{align*}
    LP(\overline{\phi}) \wedge LP(\overline{\psi}) \models c_k \ge b_k. ~~~~~~~~~~~~~~~~~~~~~~~~~~~~~~~~~~~~~~~~~~~~\qed
  \end{align*}
\end{proof}

If $\sosc$ is not satisfied, then the recursive
call to $\igfqc$ can generate linear equalities as stated
in Theorems \ref{the:gcase:1} and \ref{the:gcase:2} which can
make hypotheses in a Horn clause in $N_{\text{mix}}$ true, thus
deducing a mixed equality on symbols .

\begin{algorithm}[!htb]
	\label{alg:euf}
	\SetKwData{Left}{left}\SetKwData{This}{this}\SetKwData{Up}{up}
	\SetKwFunction{Union}{Union}\SetKwFunction{FindCompress}{FindCompress}
	\SetKwInOut{Input}{input}\SetKwInOut{Output}{output}
    \caption{ {\tt $\igfqceu$} \label{prob:in-out}}
	\Input{two formulas $\overline{\phi}$, $\overline{\psi}$, which are
    constructed respectively from $\phi$ and $\psi$ by flattening and purification,  \\
    $N_{\phi}$ : instances of functionality axioms for functions in $D_{\phi}$,\\
    $N_{\psi}$ : instances of functionality axioms for functions in $D_{\psi}$,\\
    where $\overline{\phi} \wedge \overline{\psi} \wedge N_{\phi} \wedge N_{\psi} \models \bot$,
}
	\Output{A formula $I$ to be a Craig interpolant for $\phi$ and $\psi$.}
	\SetAlgoLined
	\BlankLine
	Transform $\overline{\phi}\wedge N_{\phi} $ to a DNF $\vee_i \phi_i$\; \label{alg4:1}
       Transform $\overline{\psi}\wedge N_{\psi} $ to a DNF $\vee_j \psi_j$\; \label{alg4:2}
	\KwRet $I:= \vee_i \wedge_j \igfqc(\phi_i, \psi_j)$ \label{alg4:3}
\end{algorithm}

\begin{algorithm}[!htb]
	\label{alg:euf}
	\SetKwData{Left}{left}\SetKwData{This}{this}\SetKwData{Up}{up}
	\SetKwFunction{Union}{Union}\SetKwFunction{FindCompress}{FindCompress}
	\SetKwInOut{Input}{input}\SetKwInOut{Output}{output}
    \caption{ {\tt $\igfqce$ }\label{prob:in-out}}
	\Input{ $\overline{\phi}$ and  $\overline{\psi}$: two formulas, which are constructed
 respective from $\phi$ and $\psi$ by flattening and purification, \\
    $D$ : definitions for fresh variables introduced during flattening and purifying $\phi$ and $\psi$,\\
    $N$ : instances of functionality axioms for functions in $D$,\\
    where $\phi \wedge \psi \models \bot$, \\
$\overline{\phi}= f_1 \ge 0 \wedge \ldots \wedge f_{r_1} \ge 0 \wedge g_1 >0 \wedge \ldots \wedge g_{s_1} > 0 $,  \\
$\overline{\psi}= f_{r_1+1} \ge 0 \wedge \ldots \wedge f_{r} \ge 0 \wedge g_{s_1+1} >0 \wedge \ldots \wedge g_{s} > 0 $,
   where \\
$f_1, \ldots, f_{r}, g_1, \ldots, g_s$ are all CQ polynomial,\\
$f_1, \ldots, f_{r_1}, g_1, \ldots, g_{s_1} \in \RR[\xx,\yy]$, and\\
$f_{r_1+1}, \ldots, f_{r}, g_{s_1+1}, \ldots, g_{s} \in \RR[\xx,\zz]$
}
	\Output{A formula $I$ to be a Craig interpolant for $\phi$ and $\psi$}
	\SetAlgoLined
	\BlankLine
     \eIf  { $\sosc$ holds }
         { $L_1:=LP(\overline{{\phi}})$;  $L_2:=LP(\overline{{\psi}})$\; \label{alg3:7}
           separate $N$ to $N_{\phi}$, $N_{\psi}$ and $N_{mix}$\;
           $N_{\phi}, N_{\psi} := \textbf{SEPmix}(L_1, L_2, \emptyset, N_{\phi}, N_{\psi}, N_{mix})$\;
           $\overline{I} := \textbf{IGFQCEunmixed}(\overline{\phi}, \overline{\psi}, N_{\phi}, N_{\psi})$\;
      }
      { Find $\delta_1,\ldots,\delta_r \ge 0$ and an SOS
        polynomial $h$ using SDP
        s.t. $\sum_{i=1}^r \delta_i f_i +h \equiv 0$,\; \label{alg3:19}
    Construct $h_1 \in \RR[\xx,\yy]$ and $h_2 \in \RR[\xx,\zz]$ with form $(H1)$ and $(H2)$\;
    \label{alg3:20}
    $f:=\sum_{i=1}^{r_1} \delta_i f_i +h_1 =-\sum_{i=r_1}^{r} \delta_i f_i -h_2  $\; \label{alg3:22}
  Construct  $\overline{\phi'}$ and $\overline{\psi'}$ by Theorem \ref{the:gcase:1} and Theorem
    \ref{the:gcase:2} by eliminating variables due to condition
    $h_1 = h_2 = 0$\; \label{alg3:22}
     $I' := \igfqce(\overline{\phi'}, \overline{\psi'},D,N)$\;     \label{alg3:24}
     $\bar{I}:=(f>0) \vee (f \ge 0 \wedge I')$\;   }
      Obtain $I$ from $\overline{I}$\;
        \KwRet $I$
\end{algorithm}

\begin{algorithm}[!htb]
	\label{alg:euf}
	\SetKwData{Left}{left}\SetKwData{This}{this}\SetKwData{Up}{up}
	\SetKwFunction{Union}{Union}\SetKwFunction{FindCompress}{FindCompress}
	\SetKwInOut{Input}{input}\SetKwInOut{Output}{output}
    \caption{ {\tt $\textbf{SEPmix}$ }\label{prob:in-out}}
	\Input{ $L_1,L_2$: two sets of linear inequalities,\\
	$W$: a set of equalities,\\
	$N_{\phi}, N_{\psi}, N_{mix}$: three sets of instances of functionality axioms.
}
	\Output{$N_{\phi}, N_{\psi}$: s.t. $N_{mix}$ is separated into $N_{\phi}$ or $N_{\psi}$.}
	\SetAlgoLined
	\BlankLine
           \eIf {there exists $(\bigwedge_{k=1}^K c_k=b_k \rightarrow c=b) \in N_{mix}$ s.t
    $L_1 \wedge L_2 \wedge W \models \bigwedge_{k=1}^K c_k=b_k$}
            {
               \eIf { $c$ is $\phi$-local and $b$ is $\psi$-local}
               {
               for each $k \in \{ 1, \ldots, K \}$,
               $t_k^{-}, t_k^{+} := \textbf{SEP}(L_1,L_2,c_k,b_k)$\;
               $\alpha:=$ function symbol corresponding to $\bigwedge_{k=1}^K c_k=b_k \rightarrow c=b$\;
               $t:=$ fresh variable;
               $D := D \cup \{ t=f(t_1^{+}, \ldots, t_K^{+}) \}$\;
               $C_{\phi}:=\bigwedge_{k=1}^K c_k=t_k^{+} \rightarrow c=t$;
               $C_{\psi}:=\bigwedge_{k=1}^K t_k^{+}=b_k \rightarrow t=b$\;
               $N_{mix}:=N_{mix}-\{ C \}$; $N_{\phi} := N_{\phi} \cup \{ C_{\phi} \}$\;
               $N_{\psi} := N_{\psi} \cup \{ C_{\psi} \}$;
               $W:= W \cup \{ c=t,t=d \}$\;
               }
               {
                  \eIf {$c$ and $b$ are $\phi$-local}
                  {
                  $N_{mix} := N_{mix} -\{ C \}$; $N_{\phi} := N_{\phi} \cup \{ C \}$; $W:=W \cup \{ c = b \}$\;
                  }
                  {
                  $N_{mix} := N_{mix} -\{ C \}$; $N_{\phi} := N_{\phi} \cup \{ C \}$; $W:=W \cup \{ c = b \}$\;
                  }
               }
               call $\textbf{SEPmix} (L_1, L_2, W, N_{\phi}, N_{\psi}, N_{mix})$\;
            }
            {
        \KwRet $N_{\phi}$ and $N_{\psi}$\;
        }
\end{algorithm}

\begin{algorithm}[!htb]
	\label{alg:euf}
	\SetKwData{Left}{left}\SetKwData{This}{this}\SetKwData{Up}{up}
	\SetKwFunction{Union}{Union}\SetKwFunction{FindCompress}{FindCompress}
	\SetKwInOut{Input}{input}\SetKwInOut{Output}{output}
    \caption{ {\tt $\textbf{SEP}$ }\label{prob:in-out}}
	\Input{ $L_1,L_2$: two sets of linear inequalities,\\
	$c_k, b_k$: local variables from $L_1$ and $L_2$ respectively.
}
	\Output{$t^{-}, t^{+}$:  expressions over common variables of $L_1$ and $L_2$
	s.t $L_1 \models t^{-} \le c_k \le t^{+}$ and $L_2 \models t^{+} \le b_k \le t^{-}$}
	\SetAlgoLined
	\BlankLine
        rewrite $L_1$ and $L_2$ as constraints in matrix form $a - A x \ge 0$ and $b - B x \ge 0$\;
        $x_i, x_j$ in $x$ is the variable $c_k$ and $b_k$\;
        $e^{+} := \nu^{+} A + \mu^{+} B$; $e^{-} := \nu^{-} A + \mu^{-} B$\;
        $\nu^{+},\mu^{+} :=$ solution for $\nu^{+} \ge 0 \wedge \mu^{+} \ge 0 \wedge \nu^{+} a+ \mu^{+} b \le 0 \wedge
        e_i^{+}=1 \wedge e_j^{+}=-1 \wedge \bigwedge_{l \neq i,j} e_l^{+}=0$\;
        $\nu^{-},\mu^{-} :=$ solution for $\nu^{-} \ge 0 \wedge \mu^{-} \ge 0 \wedge \nu^{-} a+ \mu^{-} b \le 0 \wedge
        e_i^{-}=-1 \wedge e_j^{-}=1 \wedge \bigwedge_{l \neq i,j} e_l^{-}=0$\;
        $t^{+} := \mu^{+}Bx + x_j - \mu^{+} b$\;
        $t^{-} := \nu^{-} Ax + x_i - \nu^{-} a$\;
        \KwRet $t^{+}$ and $t^{-}$\;
\end{algorithm}

\begin{theorem} (Soundness and Completeness of $\igfqce$)
 $\igfqce$  computes an interpolant $I$ of mutually contradictory $\phi, \psi$ with CQ
 polynomial inequalities and \textit{EUF}.
\end{theorem}

\begin{proof}
Let $\phi$ and $\psi$ are two formulas satisfy the conditions of the input of the
Algorithm $\igfqce$, $D$ is the set of definitions of fresh variables introduced during
flattening and purifying $\phi$ and $\psi$, and $N$ is the set of instances of functionality
axioms for functions in $D$.

  If the condition $\sosc$ is satisfied, then from Lemma \ref{lem:linear-part},
  we could deal with $N$ just using the linear constraints in $\phi$ and $\psi$, which
  is the same as \cite{RS10}. Since $N$ is easy to be divided into three parts,
  $N_{\phi} \wedge N_{\psi} \wedge N_{\text{mix}}$. From the algorithm in \cite{RS10}, $N_{\text{mix}}$
  can be divided into two parts $N_{\phi}^{\text{mix}}$ and $N_{\psi}^{\text{mix}}$ and add them to
  $N_{\phi}$ and $N_{\psi}$, respectively. Thus, we have
  \begin{eqnarray*}
  \phi \wedge \psi \models \bot& ~\Leftrightarrow ~ & \overline{\phi} \wedge \overline{\psi} \wedge D \models \bot ~ \Leftrightarrow ~
   \overline{\phi} \wedge \overline{\psi} \wedge N_{\phi} \wedge N_{\psi} \wedge N_{\text{mix}} \models \bot \\
    & \Leftrightarrow  &\overline{\phi} \wedge N_{\phi} \wedge N_{\phi}^{\text{mix}}
   \wedge \overline{\psi} \wedge N_{\psi} \wedge N_{\psi}^{\text{mix}} \models \bot.
   \end{eqnarray*}
   The correctness of step $4$ is guaranteed by Lemma~\ref{lem:linear-part} and Theorem 8 in \cite{RS10}.
   After step $4$,  $N_{\phi}$ is replaced by $N_{\phi}\wedge N_{\phi}^{\text{mix}}$, and $N_{\psi}$ is
   replaced by $N_{\psi} \wedge N_{\psi}^{\text{mix}}$.
    An interpolant for
   $\overline{\phi} \wedge N_{\phi} \wedge N_{\phi}^{\text{mix}}$
   and $\overline{\psi} \wedge N_{\psi} \wedge N_{\psi}^{\text{mix}}$ is generated in step $5$, the correctness of
   this step is guaranteed
   by Theorem~\ref{thm:correctness-2}.
    Otherwise if the condition $\sosc$ is not satisfied, we can obtain two polynomials $h_1$ and
   $h_2$, and derive  two formulas
   $\overline{\phi'}$ and $\overline{\psi'}$. By Theorem \ref{lemma:p22}, if there is  an
   interpolant $I'$ for $\overline{\phi'}$ and $\overline{\psi'}$, then we can get an interpolant $I$
    for $\overline{\phi}$ and $\overline{\psi}$ at step $11$.
    Similar to the proof of Theorem~\ref{thm:correctness-2},
   it is easy to argue that  this reduction will terminate at  the case when $\sosc$ holds in finite steps.
   Thus, this completes the proof. \qed
\end{proof}

\begin{example}
Let two formulae $\phi$ and $\psi$ be defined as follows,
\begin{align*}
\phi:=&(f_1=-(y_1-x_1+1)^2-x_1+x_2 \ge 0) \wedge (y_2=\alpha(y_1)+1) \\
&\wedge ( g_1= -x_1^2-x_2^2-y_2^2+1 > 0),
\end{align*}
\begin{align*}
\psi:=&(f_2=-(z_1-x_2+1)^2+x_1-x_2 \ge 0)  \wedge  (z_2=\alpha(z_1)-1) \\
&\wedge  (g_2= -x_1^2-x_2^2-z_2^2+1 > 0),
\end{align*}
where $\alpha$ is an uninterpreted function. Then
\begin{align*}
\overline{\phi}:=&(f_1=-(y_1-x_1+1)^2-x_1+x_2 \ge 0) \wedge (y_2=y+1) \\
&\wedge ( g_1= -x_1^2-x_2^2-y_2^2+1 > 0),\\
\overline{\psi}:=&(f_2=-(z_1-x_2+1)^2+x_1-x_2 \ge 0)  \wedge  (z_2=z-1) \\
&\wedge  (g_2= -x_1^2-x_2^2-z_2^2+1 > 0),\\
D=(&y_1=z_1 \rightarrow y=z).
\end{align*}
The condition NSOSC is not satisfied, since
$-f_1-f_2=(y_1-x_1+1)^2+(z_1-x_2+1)^2$ is a SOS. It is easy to have
$$h_1=(y_1-x_1+1)^2~, ~~h_2=(z_1-x_2+1)^2.$$
Let $f:=f_1+h_1=-f_2-h_2=-x_1+x_2$, then it is easy to see that
$${\phi} \models f \ge0 ~,~~{\psi} \models f \le0.$$
Next we turn to find an interpolant for the following formulae
$$((\phi \wedge f>0) \vee (\phi \wedge f=0)) ~~and ~~ ((\psi \wedge -f>0) \vee (\psi \wedge f=0)).$$
Then
\begin{align}
\label{int:eq:e}
(f>0) \vee (f\ge0 \wedge I_2)
\end{align}
is an interpolant for $\phi$ and $\psi$,
where $I_2$ is an interpolant for $ \phi \wedge f=0$ and $\psi \wedge f=0$.
It is easy to see that
\begin{align*}
 \phi \wedge f=0 \models y_1=x_1-1 ~,~~ \psi \wedge f=0 \models z_1=x_2-1.
\end{align*}
Substitute then into $f_1$ in $\overline{\phi}$ and $\overline{\psi}$, we have
\begin{align*}
\overline{\phi'}=&-x_1+x_2 \ge 0 \wedge y_2=y+1 \wedge g_1>0 \wedge y_1=x_1-1,\\
\overline{\psi'}=&~~~~x_1-x_2 \ge 0 \wedge z_2=z-1 \wedge g_2>0 \wedge z_1=x_2-1.
\end{align*}
Only using the linear form in $\overline{\phi'}$ and $\overline{\psi'}$ we deduce that $y_1=z_1$ as
\begin{align*}
\overline{\phi'} \models t^{-}=x_1-1 \le y_1 \le t^{+}=x_2-1~~,~~\overline{\psi'} \models x_2-1 \le z_1 \le x_1-1.
\end{align*}
Let $t=\alpha(t)$, then separate $y_1=z_1 \rightarrow y=z$ into two parts,
\begin{align*}
y_1=t^{+} \rightarrow y=t, ~~ t^{+}=z_1 \rightarrow t=z.
\end{align*}
Add them to $\overline{\phi'}$ and $\overline{\psi'}$ respectively, we have
\begin{align*}
\overline{\phi'}_1=&-x_1+x_2 \ge 0 \wedge y_2=y+1 \wedge g_1>0 \wedge y_1=x_1-1 \wedge y_1=x_2-1 \rightarrow y=t,\\
\overline{\psi'}_1=&~~~~x_1-x_2 \ge 0 \wedge z_2=z-1 \wedge g_2>0 \wedge z_1=x_2-1 \wedge x_2-1=z_1 \rightarrow t=z.
\end{align*}
Then
\begin{align*}
\overline{\phi'}_1=&-x_1+x_2 \ge 0 \wedge y_2=y+1 \wedge g_1>0 \wedge y_1=x_1-1 \wedge \\
    & (x_2-1>y_1 \vee  y_1>x_2-1 \vee y=t),\\
\overline{\psi'}_1=&~~~~x_1-x_2 \ge 0 \wedge z_2=z-1 \wedge g_2>0 \wedge z_1=x_2-1 \wedge t=z.
\end{align*}
Thus,
\begin{align*}
\overline{\phi'}_1=&\overline{\phi'}_2\vee \overline{\phi'}_3 \vee \overline{\phi'}_4,\\
\overline{\phi'}_2=&-x_1+x_2 \ge 0 \wedge y_2=y+1 \wedge g_1>0 \wedge y_1=x_1-1 \wedge x_2-1>y_1,\\
\overline{\phi'}_3=&-x_1+x_2 \ge 0 \wedge y_2=y+1 \wedge g_1>0 \wedge y_1=x_1-1 \wedge y_1>x_2-1,\\
\overline{\phi'}_4=&-x_1+x_2 \ge 0 \wedge y_2=y+1 \wedge g_1>0 \wedge y_1=x_1-1 \wedge y=t.
\end{align*}
Since
$\overline{\phi'}_3=false$, then
$\overline{\phi'}_1=\overline{\phi'}_2\vee \overline{\phi'}_4$.
Then find interpolant
$$I(\overline{\phi'}_2,\overline{\psi'}_1),~~~~I(\overline{\phi'}_4,\overline{\psi'}_1). $$
$=$ replace by two $\ge$, like,
$y_1=x_1-1$ replace by $y_1\ge x_1-1$ and $x_1-1 \ge y_1$.

Then let $I_2=I(\overline{\phi'}_2,\overline{\psi'}_1) \vee I(\overline{\phi'}_4,\overline{\psi'}_1) $
an interpolant is found from (\ref{int:eq:e}) .
\end{example}

\section{Proven interpolant}
Since our result is obtained by numerical calculation, it can't guard the solution satisfy the constraints strictly.
Thus, we should verify the solution obtained from a $\sdp$ solver to get a proven interpolant.
In the end of section \ref{sec:sdp}, the remark \ref{remark:1} said one can use Lemma \ref{lem:split} to verify the
result obtained from some $\sdp$ solver. In this section, we illuminate how to verify the result obtained from some
$\sdp$ solver to get a proven interpolant by an example.

\begin{example}
  { \begin{align*}
\phi :&=f_1=4-(x-1)^2-4y^2 \ge 0 \wedge f_2=y- \frac{1}{2} \ge 0, \\
\psi :&=f_3=4-(x+1)^2-4y^2 \ge 0 \wedge f_4=x+2y \ge 0.
\end{align*} }
\end{example}
Constructing SOS constraints as following,
\begin{align*}
  &\lambda_1\ge 0, \lambda_2\ge 0, \lambda_3\ge 0, \lambda_4 \ge 0, \\
  &-(\lambda_1 f_1+ \lambda_2f_2+\lambda_3 f_3+ \lambda_4f_4+1) \mbox{ is a SOS polynomial}
\end{align*}
Using the $\sdp$ solver \textit{Yalmip} to solve the above constraints for $\lambda_1, \lambda_2, \lambda_3, \lambda_4$, 
take two decimal places, we obtain
\begin{align*}
  \lambda_1=3.63, \lambda_2=38.39, \lambda_3=0.33, \lambda_4=12.70.
\end{align*} 
Then we have,
\begin{align*}
  -(\lambda_1 f_1+ \lambda_2f_2+\lambda_3 f_3+ \lambda_4f_4+1)
  =3.96x^2+6.10x+15.84y^2-12.99y+6.315.
\end{align*}
Using Lemma \ref{lem:split}, we have 
{\small
\begin{align*}
  3.96x^2+6.10x+15.84y^2-12.99y+6.315
  =3.96(x+\frac{305}{396})^2+15.84(y+\frac{1299}{3168})^2+\frac{825383}{6336},
\end{align*} }
which is a SOS polynomial obviously.
Thus, $I:=\lambda_1f_1+\lambda_2f_2+1>0$, i.e.,
$-3.63X^2-14.52y^2+7.26x+38.39y-7.305>0$, is a proven interpolant for $\phi$ and $\psi$.

\section{Beyond concave quadratic polynomials}
Theoretically speaking, \emph{concave quadratic} is quite restrictive. But in practice, the results obtained above are powerful enough
to scale up the existing verification techniques of programs and hybrid systems, as all well-known abstract domains, e.g. \emph{octagon}, \emph{polyhedra},
\emph{ellipsoid}, etc. are concave quadratic, which will be further demonstrated in the case study below. Nonetheless, we now discuss how to generalize our approach
to more general formulas by allowing polynomial equalities whose polynomials may be neither concave nor quadratic
using Gr\"{o}bner basis.

\oomit{In the above sections, we give an algorithm to find an interpolant when all the constraints are concave quadratic,
which means that if there exists an equation in the constraints it must be linear.
In this section, we allow the polynomial equation join the constraints, i.e., we suppose that all inequation constraints are
the same concave quadratic, but equation constraints are normal polynomial equations without the condition to be linear.
We give a sufficient method to generate an interpolant.}

Let's start the discussion with the  following running example.
\begin{example} \label{nonCQ-exam}
   Let $G=A \wedge B$, where
   \begin{align*}
     A:~&x^2+2x+(\alpha(\beta(a))+1)^2 \leq 0 \wedge \beta(a)=2c+z \wedge\\
     &2c^2+2c+y^2+z=0 \wedge -c^2+y+2z=0,\\
     B:~&x^2-2x+(\alpha(\gamma(b))-1)^2 \leq 0 \wedge \gamma(b)=d-z \wedge\\
     &d^2+d+y^2+y+z=0 \wedge -d^2+y+2z=0,
   \end{align*}
   try to find an interpolant for $A$ and  $B$.
\end{example}

It is easy to see that there exist some constraints which are not concave quadratic, as some equations are not linear.
Thus, the interpolant generation algorithm above is not applicable directly.

For easing discussion, in what follows, we use $\ieq(S), \eq(S)$ and $\leqs(S)$ to stand for the sets of polynomials respectively from inequations,
  equations and linear equations of $S$, for any polynomial formula $S$. E.g.,
  in Example \ref{nonCQ-exam}, we have
  \begin{align*}
    \ieq(A)&= \{ x^2+2x+(\alpha(\beta(a))+1)^2 \},\\
    \eq(A)&= \{\beta(a)-2c-z, 2c^2+2c+y^2+z, -c^2+y+2z\},\\
    \leqs(A)&=\{\beta(a)-2c-z \}.
  \end{align*}

\oomit{
\begin{problem} \label{nonCQ-problem}
Let $A(\xx,\zz)$ and $B(\yy,\zz)$ be
 \begin{align}
  A\, :\, &f_1(\xxx,\zz)\ge 0 \wedge \ldots \wedge f_{r_1}(\xxx,\zz) \ge 0
  \wedge g_1(\xxx,\zz)> 0 \wedge \ldots \wedge g_{s_1}(\xxx,\zz) > 0 \nonumber\\
 \, &\wedge h_1(\xxx,\zz)= 0 \wedge \ldots \wedge h_{p_1}(\xxx,\zz) = 0, \\
  B \, :\, &f_{r_1+1}(\yy,\zz)\ge 0 \wedge \ldots \wedge f_{r}(\yy,\zz) \ge 0
  \wedge g_{s_1+1}(\yy,\zz)> 0 \wedge \ldots \wedge g_{s}(\yy,\zz) > 0 \nonumber\\
  \,  &\wedge h_{p_1+1}(\yy,\zz)= 0 \wedge \ldots \wedge h_{p}(\yy,\zz) = 0,
\end{align}
where $f_1, \ldots, f_r$ and $g_1, \ldots, g_s$ are concave quadratic polynomials,
$h_1, \ldots, h_t$ are general polynomials, unnecessary to be concave quadratic, and
\begin{align}
  A(\xxx,\zz) \wedge B(\yy,\zz) \models \bot,
\end{align}
try to find an interpolant for $A(\xxx,\zz)$ and $B(\yy,\zz)$.
\end{problem}

\begin{note}

  Let's denote $\ieq(S), \eq(S)$ and $\leqs(S)$ to be the set of all inequations in $S$,
  all equations in $S$ and all linear equations in $S$, where $S$ is a formula. E.g.,
  formula $A$ in Example \ref{nonCQ-exam}, we have
  \begin{align*}
    \ieq(A)&= \{ x^2+2x+(\alpha(\beta(a))+1)^2 \},\\
    \eq(A)&= \{\beta(a)-2c-z, 2c^2+2c+y^2+z=0, -c^2+y+2z\},\\
    \leqs(A)&=\{\beta(a)-2c-z \}.
  \end{align*}
\end{note} }

In the following, we will use Example \ref{nonCQ-exam} as a running example to explain the basic idea how to
apply Gr\"{o}bner basis method to extend our approach to more general polynomial formulas. 

Step $1$: Flatten and purify.
Similar to the concave quadratic case, we purify and
 flatten  $A$ and
 $B$  by introducing  fresh variables $a_1,a_2,b_1,b_2$, and obtain
   \begin{align*}
     A_0:~&x^2+2x+(a_2+1)^2 \leq 0 \wedge a_1=2c+z \wedge \\
      & 2c^2+2c+y^2+z=0 \wedge -c^2+y+2z=0,\\
     D_A:~&a_1=\beta(a) \wedge a_2=\alpha(a_1),\\
     B_0:~&x^2-2x+(b_2-1)^2 \leq 0 \wedge b_1=d-z \wedge  \\
       & d^2+d+y^2+y+2z=0 \wedge -d^2+y+z=0,\\
     D_B:~&b_1=\gamma(b) \wedge b_2=\alpha(b_1).
   \end{align*}

 Step $2$: { Hierarchical reasoning}.
 Obviously,  $A \wedge B$ is unsatisfiable in
 $\TQ^{ \{ \alpha,\beta,\gamma\} }$
 if and only if $A_0 \wedge B_0 \wedge N_0$ is unsatisfiable in $\TQ$,
 where $N_0$
 corresponds to the conjunction of Horn clauses constructed  from $D_A \wedge D_B$ using
 the axioms of uninterpreted functions (see the following table).
{\small  \begin{center}
 \begin{tabular}{c|c|c}\hline
 D & $G_0$ & $N_0$ \\\hline
 $D_A:~a_1=\beta(a) \wedge$ & $A_0:~x^2+2x+(a_2+1)^2 \leq 0 \wedge
 a_1=2c+z \wedge$ & \\
   \quad ~~~~ $a_2=\alpha(a_1)$  & $2c^2+2c+y^2+z=0 \wedge -c^2+y+2z=0$ & $N_0: b_1=a_1 \rightarrow
 b_2=a_2$\\
    & & \\
    $D_B:~b_1=\gamma(b) \wedge$ & $B_0:~x^2-2x+(b_2-1)^2 \leq 0
 \wedge b_1=d-z \wedge$ & \\
    \quad ~~~~ $b_2=\alpha(b_1)$ & $d^2+d+y^2+y+2z=0 \wedge -d^2+y+z=0$ & \\
 \end{tabular}
 \end{center} }

 \medskip
 To prove $A_0 \wedge B_0 \wedge N_0 \models \bot$, 
 we compute the Grobner basis of  $\mathbb{G}$ of $\eq(A_0) \cup \eq(B_0)$ under the order 
 $c\succ d \succ y\succ z \succ a_1 \succeq b_1$, and have $a_1-b_1 \in \mathbb{G}$. That is,
 $A_0 \wedge B_0 \models a_1=b_1$.
 Thus, $A_0 \wedge B_0 \wedge N_0$ entails
 \begin{align*}
   a_2=b_2 \wedge x^2+2x+(a_2+1)^2 \leq 0 \wedge x^2-2x+(b_2-1)^2 \leq 0.
 \end{align*}
This implies 
 $$2x^2+a_2^2+b_2^2+2\leq0,$$
 which is obviously unsatisfiable in $\QQ$.

Step $2$ gives a proof of $A \wedge B \models \bot$.
In order to find an interpolant for $A$ and $B$, we need to divide $N_0$ into two parts,
$A$-part and $B$-part, i.e.,  to find a term $t$ only with common symbols, such that
\begin{align*}
  A_0 \models a_1=t   ~~~B_0 \models b_1=t.
\end{align*}
Then we can choose a new variable $\alpha_t=\alpha(t)$ to be a common variable, since the term $t$ and
the function $\alpha$ both are common. Thus $N_0$ can be divided into two parts as follows,
\begin{align*}
  a_2=\alpha_t \wedge b_2=\alpha_t.
\end{align*}
Finally,  if we can find an interpolant $I(x,y,z,\alpha_t)$ for
\begin{align*}
  (\ieq(A_0) \wedge \leqs(A_0) \wedge a_2=\alpha_t) \wedge (\ieq(A_0) \wedge \leqs(A_0) \wedge b_2=\alpha_t),
\end{align*} using Algorithm $\igfqc$,
then $I(x,y,z,\alpha(t))$ will be an interpolant for $A \wedge B$.

Step $3$: Dividing $N_0$ into two parts. According to the above analysis, we need to find a witness $t$ 
such that $A_0 \models a_1=t$, $B_0 \models b_1=t$, where $t$ is an expression over the common symbols of $A$ and $B$.
Fortunately, such $t$ can be computed by Gr\"{o}bner basis method as follows:
 First, with the variable order $c \succ a_1 \succ y \succ z$, the Gr\"{o}bner basis $\mathbb{G}_1$ of $\eq(A_0)$
 is computed to be
\begin{align*}
  \mathbb{G}_1=&\{ y^4+4y^3+10y^2z+4y^2+20yz+25z^2-4y-8z, \\
  &y^2+a_1+2y+4z, y^2+2c+2y+5z \}.
\end{align*}
Thus, we have
\begin{align} \label{eq-a1}
  A_0 \models a_1=-y^2-2y-4z.
\end{align}
Simiarly, with the variable order $d \succ b_1 \succ y \succ z$, the Gr\"{o}bner basis $\mathbb{G}_2$ of $\eq(B_0)$
is computed to be
\begin{align*}
  \mathbb{G}_2=&\{ y^4+4y^3+6y^2z+4y^2+12yz+9z^2-y-z, \\
  &y^2+b_1+2y+4z, y^2+d+2y+3z \}.
\end{align*}
Thus, we have
\begin{align} \label{eq-b1}
  B_0 \models b_1=-y^2-2y-4z.
\end{align}
Whence, $t=-y^2-2y-4z$ is the witness.
Let $\alpha_t=\alpha(-y^2-2y-4z)$, which is an expression constructed from the common symbols of $A$ and $B$.

Next, find an interpolant for following formula
\begin{align*}
  (\ieq(A_0) \wedge \leqs(A_0) \wedge a_2=\alpha_t) \wedge (\ieq(B_0) \wedge \leqs(B_0) \wedge b_2=\alpha_t).
\end{align*}
Using $\igfqc$, we obtain an interpolant for the above formula as
\begin{align*}
  I(x,y,z,\alpha_t)=x^2+2x+(\alpha_t+1)\le 0.
\end{align*}
Thus, $x^2+2x+(\alpha(-y^2-2y-4z)+1)\le 0$ is an interpolant for $A \wedge B$.


\begin{problem} \label{nonCQ-problem}
Generally, let $A(\xx,\zz)$ and $B(\yy,\zz)$ be
 \begin{align}
  A\, :\, &f_1(\xxx,\zz)\ge 0 \wedge \ldots \wedge f_{r_1}(\xxx,\zz) \ge 0
  \wedge g_1(\xxx,\zz)> 0 \wedge \ldots \wedge g_{s_1}(\xxx,\zz) > 0 \nonumber\\
 \, &\wedge h_1(\xxx,\zz)= 0 \wedge \ldots \wedge h_{p_1}(\xxx,\zz) = 0, \\
  B \, :\, &f_{r_1+1}(\yy,\zz)\ge 0 \wedge \ldots \wedge f_{r}(\yy,\zz) \ge 0
  \wedge g_{s_1+1}(\yy,\zz)> 0 \wedge \ldots \wedge g_{s}(\yy,\zz) > 0 \nonumber\\
  \,  &\wedge h_{p_1+1}(\yy,\zz)= 0 \wedge \ldots \wedge h_{p}(\yy,\zz) = 0,
\end{align}
where $f_1, \ldots, f_r$ and $g_1, \ldots, g_s$ are concave quadratic polynomials,
$h_1, \ldots, h_t$ are general polynomials, unnecessary to be concave quadratic, and
\begin{align}
  A(\xxx,\zz) \wedge B(\yy,\zz) \models \bot,
\end{align}
try to find an interpolant for $A(\xxx,\zz)$ and $B(\yy,\zz)$.
\end{problem}

According to the above discussion, Problem~\ref{nonCQ-problem} can be solved by Algorithm~\ref{ag:nonCQ} below. 
\begin{algorithm}[!htb] \label{ag:nonCQ}
	\label{alg:int}
	\SetKwData{Left}{left}\SetKwData{This}{this}\SetKwData{Up}{up}
	\SetKwFunction{Union}{Union}\SetKwFunction{FindCompress}{FindCompress}
	\SetKwInOut{Input}{input}\SetKwInOut{Output}{output}
    \caption{ {\tt $\igfqc$ }\label{prob:in-out}}
	\Input{Two formulae $A$, $B$ as Problem \ref{nonCQ-problem} with $A \wedge B \models \bot$}
	\Output{An formula $I$ to be a Craig interpolant for $A$ and $B$}
	\SetAlgoLined
	\BlankLine
    \textbf{Flattening, purification and hierarchical reasoning}
    obtain $A_0$, $B_0$, $N_A$, $N_B$, $N_{mix}$;\\
    $A_0:=A_0 \wedge N_A, B_0:=B_0\wedge N_B$;\\
    \While{$(\ieq(A_0) \wedge \leqs(A_0)) \wedge (\ieq(B_0)\wedge \leqs(B_0)) \not\models \bot$}
    {
    \If{$N_{mix}=\emptyset$}{\textbf{break}}
    Choose a formula $a_1=b_1 \rightarrow a_2=b_2 \in N_{mix}$ corresponding to function $\alpha$;\\
    $N_{mix}:=N_{mix}\setminus \{ a_1=b_1 \rightarrow a_2=b_2\}$;\\
    Computing Grobner basis $\mathbb{G}_1$ for $\eq(A_0)$ under purely dictionary ordering with
    some variable ordering that other local variable $\succ a_1 \succ$ common variable; \\
    Computing Grobner basis $\mathbb{G}_2$ for $\eq(B_0)$ under purely dictionary ordering with
    some variable ordering that other local variable $\succ b_1 \succ$ common variable; \\
    \If{there exists a expression $t$ with common variable s.t. $a_1 \in \mathbb{G}_1\wedge b_1 \in \mathbb{G}_2$}
    {introduce a new variable $\alpha_t=\alpha(t)$ as a common variable;
    $A_0:=A_0 \wedge a_2=\alpha_t, B_0:=B_0 \wedge b_2=\alpha_t$}
    }
    \If{$(\ieq(A_0) \wedge \leqs(A_0)) \wedge (\ieq(B_0)\wedge \leqs(B_0)) \models \bot$}
    {
    Using $\igfqc$ to obtain an interpolant $I_0$ for above formula;\\
    Obtain an interpolant $I$ for $A \wedge B$ from $I_0$;\\
    \KwRet $I$ }
    \Else{\KwRet Fail}
\end{algorithm}

\section{Implementation and experimental results}
We have implemented the presented algorithms in \textit{Mathematica} to synthesize interpolation for concave quadratic polynomial inequalities as well as their combination with \textit{EUF}. To deal with SOS solving and semi-definite programming, the Matlab-based optimization tool \textit{Yalmip} \cite{Yalmip} and the SDP solver \textit{SDPT3} \cite{SDPT3} are invoked. In what follows we demonstrate our approach by some examples, which have been evaluated on a 64-bit Linux computer with a 2.93GHz Intel Core-i7 processor and 4GB of RAM.

\begin{example} \label{exp:4}
Consider the example: 
\begin{eqnarray*}
\phi := (f_1 \ge 0) \wedge (f_2 \ge0) \wedge (g_1 >0),\quad
\psi := (f_3 \ge 0). \quad
\phi \wedge \psi \models \bot.
\end{eqnarray*}
where $f_1 = x_1, f_2 = x_2,f_3= -x_1^2-x_2^2 -2x_2-z^2, g_1= -x_1^2+2 x_1 - x_2^2 + 2 x_2 - y^2$.
\vspace{-1mm}

The interpolant returned after $0.394$ s is
\begin{eqnarray*}
I:= \frac{1}{2}x_1^2+\frac{1}{2}x_2^2+2x_2 > 0
\end{eqnarray*}
\end{example}

\begin{example} \label{exp:5}
Consider the unsatisfiable conjunction $\phi \wedge \psi$:
\begin{eqnarray*}
\phi := f_1 \ge 0 \wedge f_2 \ge 0 \wedge f_3 \ge 0 \wedge g_1>0, \quad
\psi := f_4 \ge 0 \wedge f_5 \ge 0  \wedge f_6 \ge 0 \wedge g_2 >0 .
\end{eqnarray*}
where
 $f_1  =  -y_1+x_1-2$,
 $f_2 =  -y_1^2-x_1^2+2x_1 y_1 -2y_1+2x_1$,
$f_3 =  -y_2^2-y_1^2-x_2^2 -4y_1+2x_2-4$,
$f_4  =  -z_1+2x_2+1$,
$f_5  =  -z_1^2-4x_2^2+4x_2 z_1 +3z_1-6x_2-2$,
$f_6  =  -z_2^2-x_1^2-x_2^2+2x_1+z_1-2x_2-1$,
 $g_1  =  2x_2-x_1-1$,
 $g_2  =  2x_1-x_2-1$.
\vspace{2mm}

The condition NSOSC does not hold, since
$$-(2f_1 + f_2) = (y_1 -x_1 +2)^2 \textrm{ is a sum of square}.$$
Then we have $h=(y_1 -x_1 +2)^2$, and
\begin{align*}
h_1=h=(y_1 -x_1 +2)^2, \quad h_2=0.
\end{align*}
Let $f=2f_1+f_2+h_1=0$. Then construct $\phi'$ by setting $y_1=x_1-2$ in $\phi$, $\psi'$ is $\psi$. That is
\begin{align*}
\phi':=0\ge0 \wedge 0 \ge 0 \wedge -y_2^2-x_1^2-x_2^2+2x_2 \ge 0 \wedge g_1>0, \quad \psi':=\psi.
\end{align*}
Then the interpolation for $\phi$ and $\psi$ is reduced as
\begin{align*}
I(\phi,\psi)=(f> 0) \vee (f=0 \wedge I(\phi', \psi')) = I(\phi',\psi').
\end{align*}

For $\phi'$ and $\psi'$, the condition NSOSC is still  unsatisfied, since $-f_4-f_5 = (z_1-2x_2-1)^2$ is an SOS.
Then we have $h=h_2=(z_1-2x_2-1)^2$, $h_1=0$, and thus $f=0$.
\begin{align*}
\phi''=\phi', \quad \psi''=0\ge0 \wedge 0 \ge 0 \wedge -z_2^2-x_1^2 - x_2^2+2x_1\ge0 \wedge
g_2 >0.
\end{align*}
The interpolation for $\phi'$ and $\psi'$ is further reduced by $ I(\phi',\psi')=I(\phi'',\psi'')$, where
\begin{align*}
\phi'':=(f_1'=-y_2^2-x_1^2-x_2^2+2x_2\ge0) \wedge 2x_2-x_1-1>0,\\
\psi'':=(f_2'=-z_2^2-x_1^2 - x_2^2+2x_1\ge0) \wedge
2x_1-x_2-1 >0.
\end{align*}

Here the condition NSOSC holds for $\phi''$ and $\psi''$, then by SDP we find $\lambda_1=\lambda_2=0.25, \eta_0=0, \eta_1=\eta_2=0.5$ and SOS polynomials $h_1=0.25*((x_1-1)^2+(x2-1)^2+y_2^2)$ and $h_2=0.25*((x_1-1)^2+(x_2-1)^2+z_2^2)$ such that
$\lambda_1 f_1'+\lambda_2 f_2' + \eta_0 + \eta_1 g_1 + \eta_2 g_2 +h_1 +h_2 \equiv 0$ and
$\eta_0+\eta_1+\eta_2=1$.
For $\eta_0+\eta_1=0.5>0$, the interpolant returned after $2.089$ s is $f>0$, i.e. $I:= -x_1 + x_2 > 0$.
\end{example}
\begin{example} \label{exp:6}
Consider the example: 
\begin{align*}
\phi:=&(f_1=-(y_1-x_1+1)^2-x_1+x_2 \ge 0) \wedge (y_2=\alpha(y_1)+1) \\
&\wedge ( g_1= -x_1^2-x_2^2-y_2^2+1 > 0), \\
\psi:=&(f_2=-(z_1-x_2+1)^2+x_1-x_2 \ge 0)  \wedge  (z_2=\alpha(z_1)-1) \\
&\wedge  (g_2= -x_1^2-x_2^2-z_2^2+1 > 0).
\end{align*}
where $\alpha$ is an uninterpreted function. It takes $0.369$ s in our approach to reduce the problem to find an interpolant as $I(\overline{\phi'}_2,\overline{\psi'}_1) \vee (\overline{\phi'}_4,\overline{\psi'}_1)$, and another $2.029$ s to give the final interpolant as
\begin{eqnarray*}
I:= (-x_1 + x_2 > 0) \vee (\frac{1}{4}(-4\alpha(x_2-1)-x_1^2-x_2^2)>0)
\end{eqnarray*}
\end{example}
\begin{example} \label{exp:7}
Let two formulae $\phi$ and $\psi$ be defined as
\begin{align*}
\phi:=&(f_1=4-x^2-y^2 \ge 0) \wedge f_2=y \ge 0 \wedge ( g= x+y-1 > 0), \\
\psi:=&(f_4=x \ge 0) \wedge  (f_5= 1-x^2- (y+1)^2 \ge 0).
\end{align*}

The interpolant returned after $0.532$ s is $I:= \frac{1}{2}(x^2+y^2+4y)>0$ \footnote{In order to give a more objective comparison of performance with the approach proposed in \cite{DXZ13}, we skip over line 1 in the  previous algorithm $\igfqc$.}.
\oomit{While AiSat gives
\begin{small}
\begin{align*}
I':= &-0.0732 x^4+0.1091 x^3 y^2+199.272 x^3 y+274818 x^3-0.0001 x^2 y^3-0.079 x^2 y^2 \\
&-0.1512 x^2 y-0.9803 x^2+0.1091 x y^4+199.491 x y^3+275217 x y^2+549634 x y \\
&+2.0074 x-0.0001 y^5-0.0056 y^4-0.0038 y^3-0.9805 y^2+2.0326 y-1.5 > 0
\end{align*}
\end{small}}
\end{example}
\begin{example} \label{exp:8}
This is a linear interpolation problem adapted from \cite{RS10}. Consider the unsatisfiable conjunction $\phi \wedge \psi$:
\begin{eqnarray*}
\phi := z-x \ge 0 \wedge x-y \ge 0 \wedge -z > 0, \quad
\psi := x+y \ge 0 \wedge -y \ge 0 .
\end{eqnarray*}

It takes 0.250 s for our approach to give an interpolant as $I:= - 0.8x - 0.2y > 0$.
\end{example}
\begin{example} \label{exp:9}
Consider another linear interpolation problem combined with \textit{EUF}:
\begin{eqnarray*}
\phi := f(x) \ge 0 \wedge x-y \ge 0 \wedge y-x \ge 0, \quad
\psi := -f(y) > 0 .
\end{eqnarray*}
The interpolant returned after 0.236 s is $I:= f(y) \ge 0$.
\end{example}
\begin{example} \label{exp:10}
Consider two formulas $A$ and $B$ with $A \wedge B \models \bot$, where\\
   \begin{minipage}{0.6\textwidth}
		\begin{align*}
		A :=& -{x_1}^2 + 4 x_1 +x_2 - 4 \ge 0 \wedge \\
		       & -x_1 -x_2 +3 -y^2 > 0,\\
		B :=& \mathbf{-3 {x_1}^2 - {x_2}^2 + 1 \ge 0} \wedge x_2 - z^2 \ge 0.
		\end{align*}
		Note that a concave quadratic polynomial (the bold one) from the \textit{ellipsoid} domain is involved in $B$. It takes 0.388 s using our approach to give an interpolant as $ I:=-3 + 2 x_1 + {x_1}^2 + \frac{1}{2} {x_2}^2 > 0.$ An intuitive description of the interpolant is as the purple curve in the right figure, which separates $A$ and $B$ in the panel of common variables $x_1$ and $x_2$.
    \end{minipage}
    \begin{minipage}{0.4\textwidth}
  		\centering
  		\includegraphics[width=\linewidth]{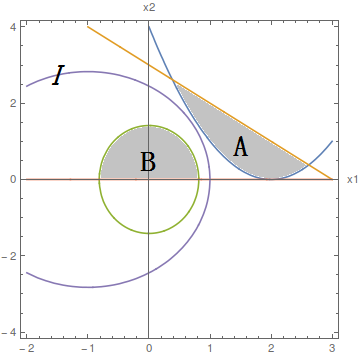}
    \end{minipage}
\end{example}

\begin{example} \label{exp:11}
Consider two formulas $\phi$ and $\psi$ both are defined by an ellipse joint a half-plane:
{\small \begin{align*}
\phi :=4-(x-1)^2-4y^2 \ge 0 \wedge y- \frac{1}{2} \ge 0,~~
\psi :=4-(x+1)^2-4y^2 \ge 0 \wedge x+2y \ge 0.
\end{align*} }
The interpolant returned after 0.248 s is $I:=-3.63x^2-14.52y^2+7.26x+38.39y-7.305  > 0$.
\end{example}

\begin{example} \label{exp:12}
Consider two formulas $\phi$ and $\psi$ both are defined by an octagon joint a half-plane:
{\small   \begin{align*}
  \phi &:= -3 \le x \le 1 \wedge -2 \le y \le 2 \wedge -4 \le x-y \le 2
  \wedge -4 \le x+y \le 2 \wedge x+2y+1\le 0, \\
  \psi &:= -1 \le x \le 3 \wedge -2 \le y \le 2 \wedge -2 \le x-y \le 4
  \wedge -2 \le x+y \le 4 \wedge 2x-5y+6\le 0.
  \end{align*} }
  The interpolant returned after 0.225 s is $I:=-13.42x-29.23y-1.7  > 0$.
\end{example}

\begin{example} \label{exp:13}
Consider two formulas $\phi$ and $\psi$ both are defined by an octagon joint a half-plane:
{\small   \begin{align*}
  \phi &:= 2 \le x \le 7 \wedge 0 \le y \le 3 \wedge 0 \le x-y \le 6
  \wedge 3 \le x+y \le 9 \wedge 23-3x-8y\le 0, \\
  \psi &:= 0 \le x \le 5 \wedge 2 \le y \le 5 \wedge -4 \le x-y \le 2
  \wedge 3 \le x+y \le 9 \wedge y-3x-2\le 0.
  \end{align*} }
  The interpolant returned after 0.225 s is $I:=12.3x-7.77y+4.12  > 0$.
\end{example}

\vspace*{-2mm}
\begin{table}
\begin{center}
\begin{tabular}{llp{1.9cm}<{\centering}p{1.4cm}<{\centering}p{1.4cm}<{\centering}p{1.4cm}<{\centering}p{1.9cm}<{\centering}}
  \toprule
  \multirow{2}{*}{\begin{minipage}{1.7cm}
      Example
    \end{minipage}}
  & \multirow{2}{*}{\begin{minipage}{1.5cm}
      Type
    \end{minipage}}
  & \multicolumn{5}{c}{Time (sec)}\\
  \cmidrule(lr){3-7}
  & & \textsc{CLP-Prover} & \textsc{Foci} & \textsc{CSIsat} & AiSat
  & Our Approach\\ \hline
  Example \ref{exp:4} & NLA     & -- & -- & -- & -- & 0.394 \\
  Example \ref{exp:5} & NLA     & -- & -- & -- & -- & 2.089 \\
  Example \ref{exp:6} & NLA+\textit{EUF} & -- & -- & -- & -- & 2.398 \\
  Example \ref{exp:7} & NLA     & -- & -- & -- & 0.023 & 0.532 \\
  Example \ref{exp:8} & LA      & 0.023 & $\times$ & 0.003 & -- & 0.250 \\
  Example \ref{exp:9} & LA+\textit{EUF}  & 0.025 & 0.006 & 0.007 & -- & 0.236 \\
  Example \ref{exp:10} & Ellipsoid & -- & -- & -- & -- & 0.388 \\
  Example \ref{exp:11} & Ellipsoid2 & -- & -- & -- & 0.013 & 0.248 \\
  Example \ref{exp:12} & Octagon1 & 0.059 & $\times$ & 0.004 & 0.021 & 0.225 \\
  Example \ref{exp:13} & Octagon2 & 0.065 & $\times$ & 0.004 & 0.122 & 0.216 \\
  \bottomrule
\end{tabular}\\
-- means that the interpolant generation fails, and $\times$ specifies a particularly wrong answer.
\vspace*{0.1in}
\caption{Evaluation results of the presented examples}\label{tab1}
\vspace*{-10mm}
\end{center}
\end{table}

%
The experimental evaluation on the above examples is illustrated in Table~\ref{tab1}, where we have also compared on the same platform with the performances of AiSat, a tool for nonlinear interpolant generation proposed in \cite{DXZ13}, as well as three publicly available interpolation procedures for linear-arithmetic cases, i.e. Rybalchenko's tool \textsc{CLP-Prover}) in \cite{RS10}, McMillan's procedure \textsc{Foci} in \cite{mcmillan05}, and Beyer's tool \textsc{CSIsat} in \cite{CSIsat}. Table \ref{tab1} shows that our approach can successfully solve all the examples and it is especially the completeness that makes it an extraordinary competitive candidate for synthesizing interpolation. Besides, \textsc{CLP-Prover}, \textsc{Foci}, and \textsc{CSIsat} can handle only linear-arithmetic expressions with an efficient optimization (and thus the performances in linear cases are better than our raw implementation). As for AiSat, a rather limited set of applications is acceptable because of the weakness of tackling local variables, and whether an interpolant can be found or not depends on a pre-specified total degree. In \cite{DXZ13}, not only all the
constraints in formula $\phi$ should be considered but also some of their products, for instance,
$f_1,f_2,f_3 \ge0$ are three constraints in  $\phi$, then four constraints $f_1f_2,f_1f_3,f_2f_3,f_1f_2f_3 \ge0$ are added in $\phi$.

Table~\ref{tab1} indicates the efficiency of our tool is lower than any of other tools whenever a considered example is
solvable by both. This is mainly because our tool is implemented in  \textit{Mathematica}, and therefore have to invoke some SDP solvers with low efficiency. As a future work, we plan to re-implement the tool using C, thus we can call SDP solver CSDP which is
much more efficient. Once a considered problem is linear, an existing interpolation procedure  will be invoked directly, thus,
 SDP solver is not needed.

\oomit{
The experimental evaluation on the above motivating examples is illustrated in table \ref{tab1}, where we have also compared on the same platform with the performances of AiSat, a tool for nonlinear interpolant generation proposed in \cite{DXZ13}, as well as three publicly available interpolation procedures for linear-arithmetic cases, i.e. Rybalchenko's tool \textsc{CLP-Prover}) in \cite{RS10}, McMillan's procedure \textsc{Foci} in \cite{mcmillan05}, and Beyer's tool \textsc{CSIsat} in \cite{CSIsat}. Table \ref{tab1} shows that our approach can successfully solve all the examples and it is especially the completeness that makes it an extraordinary competitive candidate for synthesizing interpolation. Besides, \textsc{CLP-Prover}, \textsc{Foci}, and \textsc{CSIsat} can handle only linear-arithmetic expressions with an efficient optimization. As for AiSat, a rather limited set of applications is acceptable because of the weakness of tackling local variables.}

\section{Conclusion}

The paper proposes a polynomial time algorithm for generating
interpolants from mutually contradictory conjunctions of concave
quadratic polynomial inequalities over the reals. Under a
technical condition that if no nonpositive constant combination
of nonstrict inequalities is a sum of squares polynomials, then
such an interpolant can be generated essentially using
the linearization of quadratic polynomials. Otherwise, if this
condition is not satisified, then the algorithm is recursively
called on smaller problems after deducing linear equalities
relating variables. The resulting interpolant is a disjunction of
conjunction of polynomial inequalities.

Using the hierarchical calculus framework proposed in
\cite{SSLMCS2008}, we give an interpolation algorithm for the
combined quantifier-free theory of concave quadratic polynomial
inequalities and equality over uninterpreted function
symbols. The combination algorithm is patterned after a
combination algorithm for the combined theory of linear
inequalities and equality over uninterpreted function symbols.

In addition, we also discuss how to
extend our approach to formulas with polynomial equalities whose polynomials may be neither concave nor quadratic
using Gr\"{o}bner basis.

The proposed approach is applicable to all existing abstract domains like
\emph{octagon}, \emph{polyhedra}, \emph{ellipsoid} and so on, therefore it can be used to improve
the scalability of existing verification techniques for programs and hybrid systems.

An interesting issue raised by the proposed framework for dealing
with nonlinear polynomial inequalities is the extent to which their linearization
with some additional conditions
on the coefficients (such as concavity for quadratic polynomials)
can be exploited. We are also investigating how results reported
for nonlinear polynomial inequalities based on positive
nullstellensatz \cite{Stengle} in \cite{DXZ13} and
the Archimedian condition on variables, implying that every
variable ranged over a bounded interval, can be exploited in the
proposed framework for dealing with polynomial inequalities.












\end{document}